\documentclass[arxiv, usenatbib]{iupartex}
\usepackage{newtxtext,newtxmath}
\usepackage[T1]{fontenc}
\usepackage{ae,aecompl}

\DeclareUnicodeCharacter{2212}{-} 

\usepackage{multicol}   
\usepackage{bm}		    
\usepackage{pdflscape}	

\title[An Up to date line list for spectroscopic analysis of F and G Stars]{An Updated Line List for Spectroscopic Investigation of G Stars- I: Redetermination of the Abundances in the Solar Photosphere}

\author[\c{S}ahin et. al]{T. \c{S}ahin$^{1\cc}$\orcid{0000-0002-0296-233X},
M. Marı\c{s}mak$^{1}$\orcid{0000-0002-9397-2778},
N. \c{C}ınar$^{1}$\orcid{0000-0002-5155-9280},
and S. Bilir$^{2}$\orcid{0000-0003-3510-1509}
\affsep \\
$^1$ Akdeniz University, Faculty of Science, Department of Space Sciences and Technologies 07058, Antalya, Turkey\\
$^2$ Istanbul University, Faculty of Science, Department of Astronomy and Space Sciences 34119, Beyazıt, Istanbul, Turkey
}
\corres{Timur \c{S}ahin}{timursahin@akdeniz.edu.tr}

\pubyear{2023}

\doiheader{XXXXXXX/PAR.20XX.00000}
\date{
	\pSubmit{00.00.0000} 
	\pRevReq{00.00.0000}
	\pLastRevRec{00.00.0000}
	\pAccept{00.00.0000}
	\pPubOnl{00.00.0000}
}
\volume{0}
\volnumber{1}

\begin{document}
\label{firstpage}
\pagerange{\pageref*{firstpage}--\pageref*{lastpage}}
\maketitle

\begin{abstract}
We propose a line list that may be useful for the abundance analysis of G-type stars in the wavelength range 4080 -- 6780 \AA. It is expected that the line list will be useful for surveys/libraries with overlapping spectral regions (e.g. {\sc ELODIE}/{\sc SOPHIE} libraries, UVES-580 setting of {\it Gaia}-ESO), and in particular for the analysis of F- and G-type stars in general. The atomic data are supplemented by detailed references to the sources. We estimated the Solar abundances using stellar lines and the high-resolution Kitt Peak National Observatory (KPNO) spectra of the Sun to determine the uncertainty in the $\log gf$ values. By undertaking a systematic search that makes use of the lower excitation potential and $gf$-values and using revised multiplet table as an initial guide, we identified 363 lines of 24 species that have accurate $gf$-values and are free of blends in the spectra of the Sun and a Solar analogue star, HD \,218209 (G6V), for which accurate and up-to-date abundances were obtained from both {\sc ELODIE} and {\sc PolarBASE} spectra of the star. For the common lines with the {\it Gaia}-ESO line list v.6 provided by the {\it Gaia}-ESO collaboration, we discovered significant inconsistencies in the $gf$-values for certain lines of varying species.
\end{abstract}




\section{Introduction}

Mid-spectral type main-sequence stars (F and G-type stars) play a significant role in understanding the Galactic chemical evolution and history of Galactic structure. These stars formed in the early Milky Way or are currently forming, have main-sequence lifetimes comparable to the age of the Galaxy. F and G-type main-sequence stars have an internal structure consisting of a radiative core surrounded by a large envelope. Such structural configuration prevents heavy elements produced in the core from mixing into the stellar atmosphere. Consequently, mid-spectral type stars carry the chemical composition of the molecular cloud in which they were born. The study of the abundances of heavy elements detected in the atmospheres of F and G-type stars provides valuable insights into the Galactic chemical evolution and the history of Galactic structure, as noted by \citep{pagel1975}.

Moreover, these stars offer crucial information about the formation of different populations within the Galaxy, including the halo, thick disc, and thin disc. By analyzing the kinematics and orbital dynamics of stars that have been spectroscopically studied, one can distinguish between these different population groups in the Milky Way. In addition to providing insights into Galactic chemical evolution and the formation of the Galaxy, considering the case for pure spectroscopic analysis, for instance, the $\alpha$-element abundances determined through spectroscopic analysis of stars play a crucial role in testing the population membership of host galaxies.

The abundance of elements in a stellar spectrum, such as metallicity ([Fe/H]), can also be used to estimate the age of the star. This is because the youngest stars have comparatively higher metal abundances and metallicity, while the oldest stars have lower ones \citep{placco2021}. Therefore, pure spectroscopic analysis alone provides crucial information about the nature of host galaxies, in addition to insights into the age and chemical composition of individual stars.

The kinematics, orbital dynamics, and chemical properties of stars are not only key in understanding the chemical structure of our Galaxy and its accompanying formation scenarios but also in determining the origins of (metal-poor) young and old G-spectral type stars. In this context, the author's group is currently conducting a comprehensive study involving over 90 metal-poor G-type stars in the Solar neighbourhood. This study aims to investigate the origins of these stars by analyzing their kinematics, orbital dynamics, and chemical properties. This research will shed light on the formation processes and evolutionary pathways of metal-poor G spectral-type stars in our Galaxy.

Furthermore, the author's group has previously conducted a thorough investigation of six metal-poor F-type dwarf stars in the Solar neighbourhood, with [Fe/H] values ranging from $-$2.4 to $-$1 dex. It is worth noting that the studied metal-poor F-type dwarf stars exhibited significant changes in model atmosphere parameters (effective temperature, $T_{\rm eff}$; surface gravity, $\log g$; metallicity, [Fe/H]; microturbulence, $\xi$), as reported in the literature. This research, as published in \citet{sahin2020}, employed state-of-the-art analysis methods, including classical spectroscopy with the help of {\sc ELODIE} spectra. By combining these analysis techniques, the authors were able to determine the Galactic origin of these F-type dwarf stars in the Solar neighbourhood.

The determination of accurate metal abundances depends on the accurate determination of model parameters. Current spectroscopic sky survey programs for metal-poor late spectral stars, for example, indicate the need to determine precisely calibrated model parameters for them. These current sky survey programs include  {\it Gaia}-ESO Public Spectroscopic Survey \citep[GES;][]{gilmore2012}, GALactic Archaeology with HERMES \citep[GALAH;][]{heijmans2012, silva2015, martell2017}, the Large Sky Area Multi-Object Fiber Spectroscopic Telescope \citep[LAMOST;][]{zhao2012}, the Sloan Extension for Galactic Understanding and Exploration \citep[SEGUE;][]{yanny2009}, the Apache Point Observatory Galactic Evolution Experiment \citep[APOGEE;][]{allende2008}, and the RAdial Velocity Experiment \citep[RAVE;][]{steinmetz2006}, and demonstrate that precisely determined and calibrated model atmosphere parameters are required in different regions of the electromagnetic spectrum due to differences in adopted observational methods and techniques in these surveys, for which reliable line lists and atomic data are required.

In this study, we present an up-to-date line list for planned spectral analyses in the framework of the comprehensive study mentioned above. HD\,218209 of G6V, the most metal-rich star among 90 G type program stars ($-2.5 <{\rm  [Fe/H] (dex)}<-0.5$), was selected for the preparation of the line list which is expected to be useful for the surveys/libraries with overlapping spectral regions \cite[e.g. {\sc ELODIE} library;][]{soubiran2003} and for the analysis of F and G-type stars in general. 

This paper is organized as follows. Section 2 provides information concerning the observations. Section 3 explains the procedures used to measure and identify lines, as well as the techniques employed to determine the model parameters and conduct chemical abundance analyses of both HD\,218209 and the Sun. Section 4 is dedicated to discussing findings and their potential implications. 

\section{Observations}

For the creation of the line list in this study, high resolution ($R \approx$ 42\,000) and high signal to noise ($S/N$ = 165 at 550 nm) ELODIE spectrum (HJD\,2451184.23521; $R\approx$ 42\,000; exposure time 1800 s) of HD\,218209 was selected. The ELODIE cross-dispersed echelle spectrograph provided spectral coverage from 3900 to 6800 \AA. The spectrum was continuum normalized and wavelength calibrated, and the radial velocity (RV) was corrected by the data reduction pipeline at the telescope. Since some problems were encountered in the continuum normalization of the spectra from the library, the spectrum was renormalized.

Since high resolution ($R \approx$ 76\,000 ) and high signal to noise ($S/N$ = 140 at 550 nm) {\sc PolarBASE}\footnote{http://polarbase.irap.omp.eu} \citep{petit2014} Narval\footnote{Narval spectropolarimeter is adapted to the 2m Bernard Lyot telescope and provides high-resolution spectral and polarimetric data.} spectrum (HJD\,2456232.48238; exposure time 400 s) of the star was also available, it was used to test model parameters for the star. Prior to the line measurement process, the {\sc PolarBASE} spectrum was also renormalized and corrected for the radial velocity (RV). For RV correction, we used a Python interface and the NARVAL atomic line library containing atomic transitions from 4000 to 6800 \AA. For renormalization, we used an in-house developed interactive normalization code LIME \citep{sahin2017} in Interactive Data Language ({\sc IDL}) prior to the abundance analysis. 

\begin{figure*}
\centering
\includegraphics[width=1.0\linewidth]{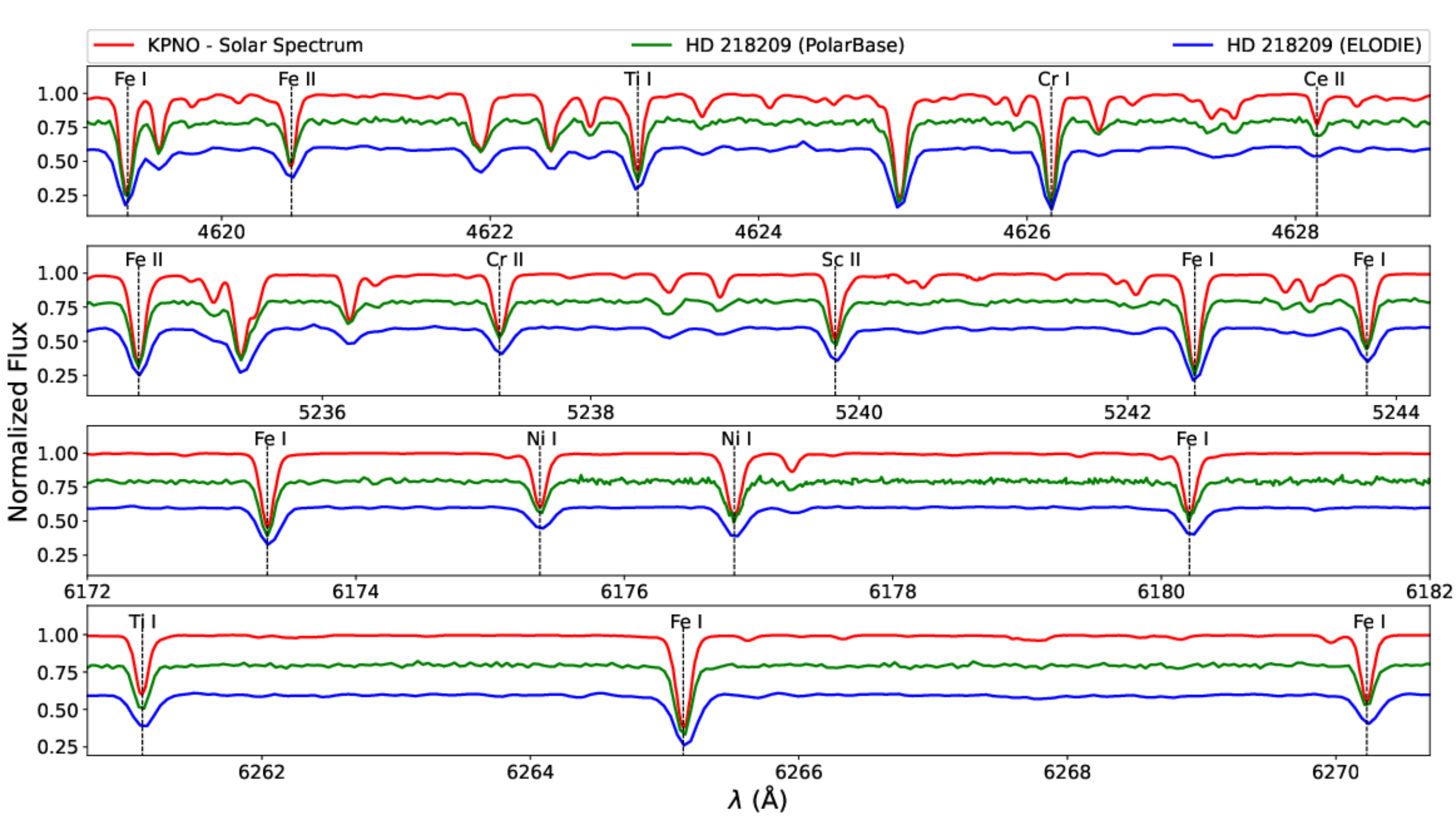}
\caption{A small region of the KPNO, {\sc PolarBASE}, and {\sc ELODIE} spectra of HD\,218209 and the Sun. Identified lines are also indicated.}
\label{fig:all_spectra}
\end{figure*}

The Solar spectrum is certainly a primary reference for stellar astrophysics and for interpreting physical processes in stars \citep{Molaro2012}. The high-resolution ($R\approx$ 400 000) spectrum of the Sun used in this study was obtained with the Kitt Peak Fourier Transform Spectrometer \citep[FTS;][]{Kurucz1984}. The character of the spectra of HD\,218209 and the Sun is displayed in Figure \ref{fig:all_spectra}.

\section{The Abundance Analysis}

We employed {\sc ATLAS9} model atmospheres \citep{Castelli2003} computed in local thermodynamic equilibrium ({\sc LTE}) with NEWODF opacities for abundance study of HD\,218209 and the Solar spectrum. Elemental abundances were computed by using the {\sc LTE} line analysis code MOOG \citep{sneden1973}\footnote{The {\sc MOOG} source code is available at http://www.as.utexas.edu/chris/moog.html}. The details of the abundance analysis and the source of the atomic data are the same as in \citet{sahin2009}, \citet{sahin2011,sahin2016} and \citet{sahin2020}. The line list, atomic data, and model parameter derivation are covered in the following subsections.

\subsection{Line List: Line Measurement, Identification and Atomic Data}

\begin{figure*}
\centering
\includegraphics[width=1.0\linewidth]{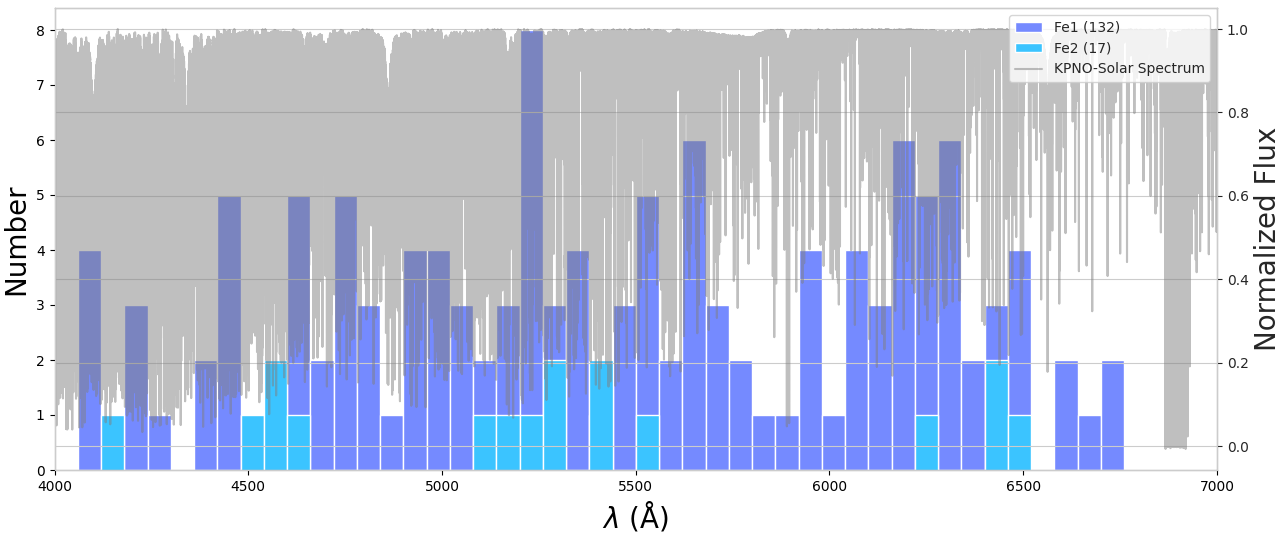}
\caption{The KPNO Solar spectrum and the number of identified lines in the respective wavelength regions (each bar indicates 50 \AA\, region of the spectrum).}
\label{fig:lines_ident}
\end{figure*}

An essential prerequisite for abundance analysis of a star is a set of reliably identified lines with reliable atomic data. Our line lists were created by a systematic search for unblended lines (useful for equivalent width--EW-- analysis technique). For line measurement from the ELODIE spectrum of HD\,218209, a systematic search for unblended lines was performed. The line centre positions were measured in several segments, each containing a portion of the spectra of 20 \AA. For this, the LIME \citep{sahin2017} code was employed. The code provides a list of possible transitions in the close neighbourhood of the measured line together with recent atomic data (e.g. Rowland Multiplet Number-RMT, $\log gf$, and lower Level Excitation Potential-LEP) that are compiled from the literature (e.g. from {\sc NIST} database). The MOORE catalogue \citep{moore1966} was configured as one of the reference atomic line libraries in the LIME code. Following the line identification step, a multiplet analysis technique was performed to assess whether the detected lines belong to the candidate element indicated by the code. Our final list covers 24 species and 363 lines over the spectrum range from about 4100 -- 6800 \AA. The identified lines have accurate $gf$-values and are free of blends in the spectra of the Sun and HD\,218209. The number of identified lines in the respective wavelength regions of the KPNO Solar spectrum is presented in Figure \ref{fig:lines_ident}. Our selection of iron lines included 132 Fe\,{\sc i} lines with excitation potentials (LEPs) ranging from 0.05 to $\approx$5 eV and 17 Fe\,{\sc ii} lines. Chosen lines of Fe\,{\sc i} and Fe\,{\sc ii} are exhibited in Table \ref{tab:lineslit_fe_lines}. In Table \ref{tab:lineslit_other_lines}, we provide the list of identified lines other than iron with the atomic data, their measured equivalent widths (obtained using the {\sc LIME} code) and computed logarithmic abundances in the Solar spectrum and in the spectrum of HD\,218209. 

Atomic data forms the basis for line diagnostics. During the calculation of element abundances, two basic atomic data stand out, namely the lower level excitation potential (LEP) and the oscillator strength ($\log gf$). Since the LEP values are measured with great accuracy, the main source of errors in abundance analysis usually comes from the $\log gf$ measurements. The sensitivity of the measurements to determine the $\log gf$ values affects the sensitivity of the spectral analysis performed. This situation may lead to incorrect calculation of model atmosphere parameters. Astrophysical $gf$-values were avoided due to the model dependence of such a procedure. We calculated Solar abundances using stellar lines to assess the uncertainty in $\log gf$-values and to minimise systematic errors. The lines were measured from the \citet{Kurucz1984} Solar flux atlas and analysed with the Solar model atmosphere from the \citet{Castelli2003} grid for $T_{\rm eff}=5790$ K, and $\log g=4.4$ cgs. Our calculations yielded a microturbulent of 0.7 km s$^{-1}$ and the Solar abundances shown in Table \ref{tab:solar_abund}. As can be noticed, the Solar abundances were successfully replicated. We use our Solar abundances to compare stellar abundances to Solar abundances. As a result, the analysis in this study is done differently in relation to the Sun. Preference for a differential approach to analysis like this will reduce the errors caused by uncertainty in oscillator strengths, the influence of spectrograph features, and departures from LTE. Effective temperature errors in the spectroscopic excitation technique can also arise from systematic errors in oscillator strength as a function of excitation potential. The same holds true for errors with equivalent widths.

\begin{table*}
\setlength{\tabcolsep}{2pt}
\small
\caption{Fe\,{\sc i} and Fe\,{\sc ii} lines. The abundances are obtained for a model of $T_{\rm eff}=5790$ K, $\log g = 4.4$ cgs, and $\xi=$ 0.66 km 
s$^{\rm −1}$.}
\label{tab:lineslit_fe_lines}
\centering
\begin{tabular}{lccccccccccccccccc}
\hline
 & & & & \multicolumn{2}{c}{Sun} & \multicolumn{2}{c}{HD\,218209}  & & & & & &\multicolumn{2}{c}{Sun} &\multicolumn{2}{c}{HD\,218209} & \\

\cline{2-4}
\cline{7-8}
\cline{11-13}
\cline{16-17}
 Spec.	&   $\lambda$ & LEP  &  $\log(gf)$ & EW  & $\log \epsilon$(X) & EW	& $\log\epsilon$(X) & RMT &Spec. & $\lambda$ & LEP  &  $\log(gf)$ & EW  & $\log\epsilon$(X) & EW  & $\log \epsilon$(X) & RMT \\
\cline{2-4}
\cline{7-8}
\cline{11-13}
\cline{16-17}
   & (\AA)  & (eV) & (dex)   &(m\AA) & (dex) &(m\AA) & (dex)&   & & (\AA)  & (eV) & (dex)   &(m\AA) & (dex)&(m\AA) & (dex) & \\
\hline
Fe\,{\sc i} & 4080.22 & 3.28 & -1.23  & 80.9   & 7.34 & 85.1   & 7.16 & 558  & Fe\,{\sc i} & 5618.64 & 4.21 & -1.28  & 49.3   & 7.47 & 34.3   & 7.04 & 1107 \\
Fe\,{\sc i} & 4082.11 & 3.42 & -1.51  & 72.7   & 7.60 & 69.8   & 7.34 & 698  & Fe\,{\sc i} & 5624.03 & 4.39 & -1.20$^{\rm *}$  & 49.2   & 7.55 & 38.9   & 7.22 & 1160 \\
Fe\,{\sc i} & 4088.56 & 3.64 & -1.50  & 52.4   & 7.43 & 43.9   & 7.09 & 906  & Fe\,{\sc i} & 5633.95 & 4.99 & -0.32  & 67.3   & 7.62 & 53.2   & 7.23 & 1314 \\
Fe\,{\sc i} & 4090.96 & 3.37 & -1.73  & 55.5   & 7.39 & 49.9   & 7.10 & 700  & Fe\,{\sc i} & 5636.71 & 3.64 & -2.56  & 19.6   & 7.51 & 13.0   & 7.18 & 868  \\
Fe\,{\sc i} & 4204.00 & 2.84 & -1.01  & 125.1  & 7.52 & 115.9  & 7.14 & 355  & Fe\,{\sc i} & 5638.27 & 4.22 & -0.84  & 75.4   & 7.54 & 61.1   & 7.12 & 1087 \\
Fe\,{\sc i} & 4207.13 & 2.83 & -1.41  & 83.5   & 7.44 & 88.1   & 7.31 & 352  & Fe\,{\sc i} & 5641.45 & 4.26 & -1.15  & 66.0   & 7.71 & 45.4   & 7.18 & 1087 \\
Fe\,{\sc i} & 4220.35 & 3.07 & -1.31  & 91.4   & 7.63 & 70.9   & 7.01 & 482  & Fe\,{\sc i} & 5662.52 & 4.18 & -0.57  & 92.4   & 7.61 & 81.1   & 7.24 & 1087 \\
Fe\,{\sc i} & 4291.47 & 0.05 & -4.08  & 92.3   & 7.51 & 87.1   & 7.12 & 3    & Fe\,{\sc i} & 5701.56 & 2.56 & -2.22  & 87.1   & 7.68 & 70.6   & 7.17 & 209  \\
Fe\,{\sc i} & 4365.90 & 2.99 & -2.25  & 49.2   & 7.44 & 38.1   & 7.03 & 415  & Fe\,{\sc i} & 5705.47 & 4.30 & -1.36  & 37.5   & 7.38 & 25.8   & 7.01 & 1087 \\
Fe\,{\sc i} & 4389.25 & 0.05 & -4.58  & 75.4   & 7.65 & 63.9   & 7.12 & 2    & Fe\,{\sc i} & 5717.84 & 4.28 & -1.10  & 63.3   & 7.62 & 53.8   & 7.29 & 1107 \\
Fe\,{\sc i} & 4432.58 & 3.57 & -1.56  & 51.4   & 7.38 & 40.2   & 6.97 & 797  & Fe\,{\sc i} & 5741.86 & 4.26 & -1.67  & 31.5   & 7.51 & 21.8   & 7.18 & 1086 \\
Fe\,{\sc i} & 4439.89 & 2.28 & -3.00  & 48.4   & 7.49 & 34.4   & 6.99 & 116  & Fe\,{\sc i} & 5778.46 & 2.59 & -3.43  & 21.5   & 7.41 & 13.6   & 7.03 & 209  \\
Fe\,{\sc i} & 4442.35 & 2.20 & -1.25  & 187.7  & 7.54 & 167.6  & 7.09 & 68   & Fe\,{\sc i} & 5806.73 & 4.61 & -1.03  & 56.4   & 7.69 & 39.7   & 7.25 & 1180 \\
Fe\,{\sc i} & 4447.14 & 2.20 & -2.73  & 66.2   & 7.66 & 64.1   & 7.45 & 69   & Fe\,{\sc i} & 5916.26 & 2.45 & -2.99  & 54.5   & 7.62 & 40.7   & 7.17 & 170  \\
Fe\,{\sc i} & 4447.73 & 2.22 & -1.34  & 171.0  & 7.54 & 155.7  & 7.11 & 68   & Fe\,{\sc i} & 5929.68 & 4.55 & -1.38  & 39.7   & 7.66 & 28.6   & 7.32 & 1176 \\
Fe\,{\sc i} & 4502.60 & 3.57 & -2.31  & 28.9   & 7.53 & 29.1   & 7.44 & 796  & Fe\,{\sc i} & 5934.67 & 3.93 & -1.12  & 76.6   & 7.47 & 61.7   & 7.05 & 982  \\
Fe\,{\sc i} & 4556.93 & 3.25 & -2.66  & 25.9   & 7.48 & 12.7   & 6.95 & 638  & Fe\,{\sc i} & 5952.73 & 3.98 & -1.39  & 59.6   & 7.50 & 53.0   & 7.22 & 959  \\
Fe\,{\sc i} & 4593.53 & 3.94 & -2.03  & 28.3   & 7.54 & 16.1   & 7.10 & 971  & Fe\,{\sc i} & 5956.71 & 0.86 & -4.61  & 52.8   & 7.62 & --     & --   & 14   \\
Fe\,{\sc i} & 4602.01 & 1.61 & -3.15  & 72.0   & 7.58 & 64.8   & 7.21 & 39   & Fe\,{\sc i} & 6027.06 & 4.07 & -1.09  & 62.7   & 7.48 & 48.8   & 7.05 & 1018 \\
Fe\,{\sc i} & 4602.95 & 1.48 & -2.22  & 122.5  & 7.51 & 116.3  & 7.13 & 39   & Fe\,{\sc i} & 6065.49 & 2.61 & -1.53  & 118.5  & 7.45 & 102.9  & 7.00 & 207  \\
Fe\,{\sc i} & 4619.30 & 3.60 & -1.08  & 84.1   & 7.45 & 73.7   & 7.06 & 821  & Fe\,{\sc i} & 6079.02 & 4.65 & -1.10  & 45.6   & 7.59 & 34.8   & 7.26 & 1176 \\
Fe\,{\sc i} & 4630.13 & 2.28 & -2.59  & 72.7   & 7.63 & 61.9   & 7.19 & 115  & Fe\,{\sc i} & 6082.72 & 2.22 & -3.57  & 35.2   & 7.52 & 28.1   & 7.21 & 64   \\
Fe\,{\sc i} & 4635.85 & 2.84 & -2.36  & 54.9   & 7.54 & 41.2   & 7.06 & 349  & Fe\,{\sc i} & 6096.67 & 3.98 & -1.88  & 37.5   & 7.55 & 27.2   & 7.22 & 959  \\
Fe\,{\sc i} & 4678.85 & 3.60 & -0.83  & 102.5  & 7.47 & 95.2   & 7.13 & 821  & Fe\,{\sc i} & 6127.91 & 4.14 & -1.40  & 48.3   & 7.52 & 37.3   & 7.17 & 1017 \\
Fe\,{\sc i} & 4704.95 & 3.69 & -1.53  & 61.5   & 7.55 & 48.6   & 7.12 & 821  & Fe\,{\sc i} & 6137.70 & 2.59 & -1.40  & 135.6  & 7.48 & 122.8  & 7.07 & 207  \\
Fe\,{\sc i} & 4728.55 & 3.65 & -1.17  & 81.3   & 7.65 & 69.3   & 7.23 & 822  & Fe\,{\sc i} & 6157.73 & 4.07 & -1.22  & 61.0   & 7.56 & 53.2   & 7.26 & 1015 \\
Fe\,{\sc i} & 4733.60 & 1.48 & -2.99  & 82.9   & 7.58 & 84.8   & 7.40 & 38   & Fe\,{\sc i} & 6165.36 & 4.14 & -1.47  & 44.6   & 7.51 & 30.7   & 7.09 & 1018 \\
Fe\,{\sc i} & 4735.85 & 4.07 & -1.32  & 64.1   & 7.79 & 57.6   & 7.50 & 1042 & Fe\,{\sc i} & 6173.34 & 2.22 & -2.88  & 68.9   & 7.61 & 58.6   & 7.22 & 62   \\
Fe\,{\sc i} & 4741.53 & 2.83 & -1.76  & 72.6   & 7.41 & 63.5   & 7.03 & 346  & Fe\,{\sc i} & 6180.21 & 2.73 & -2.65  & 53.4   & 7.52 & 47.6   & 7.25 & 269  \\
Fe\,{\sc i} & 4745.81 & 3.65 & -1.27  & 78.2   & 7.69 & 66.2   & 7.27 & 821  & Fe\,{\sc i} & 6200.32 & 2.61 & -2.44  & 71.5   & 7.59 & 62.6   & 7.23 & 207  \\
Fe\,{\sc i} & 4788.77 & 3.24 & -1.76  & 65.6   & 7.61 & 55.0   & 7.20 & 588  & Fe\,{\sc i} & 6213.44 & 2.22 & -2.48  & 82.5   & 7.50 & 75.6   & 7.18 & 62   \\
Fe\,{\sc i} & 4802.89 & 3.64 & -1.51  & 60.0   & 7.52 & 44.3   & 7.02 & 888  & Fe\,{\sc i} & 6219.29 & 2.20 & -2.43  & 87.9   & 7.54 & 79.1   & 7.18 & 62   \\
Fe\,{\sc i} & 4839.55 & 3.27 & -1.82  & 62.2   & 7.60 & 58.3   & 7.37 & 588  & Fe\,{\sc i} & 6232.65 & 3.65 & -1.22  & 81.5   & 7.60 & 71.9   & 7.25 & 816  \\
Fe\,{\sc i} & 4875.88 & 3.33 & -1.97  & 61.0   & 7.60 & 49.2   & 7.20 & 687  & Fe\,{\sc i} & 6240.65 & 2.22 & -3.17  & 47.6   & 7.40 & 39.1   & 7.06 & 64   \\
Fe\,{\sc i} & 4917.23 & 4.19 & -1.16  & 62.7   & 7.61 & 49.4   & 7.21 & 1066 & Fe\,{\sc i} & 6252.56 & 2.40 & -1.69  & 119.6  & 7.42 & 103.2  & 6.96 & 169  \\
Fe\,{\sc i} & 4918.02 & 4.23 & -1.34  & 53.0   & 7.63 & 45.5   & 7.35 & 1070 & Fe\,{\sc i} & 6265.14 & 2.18 & -2.55  & 86.2   & 7.60 & 82.5   & 7.33 & 62   \\
Fe\,{\sc i} & 4924.78 & 2.28 & -2.11  & 92.8   & 7.52 & 85.2   & 7.15 & 114  & Fe\,{\sc i} & 6270.23 & 2.86 & -2.61  & 53.0   & 7.58 & 43.1   & 7.23 & 342  \\
Fe\,{\sc i} & 4939.69 & 0.86 & -3.34  & 99.7   & 7.58 & 95.8   & 7.24 & 16   & Fe\,{\sc i} & 6297.80 & 2.22 & -2.74  & 73.6   & 7.56 & 67.1   & 7.26 & 62   \\
Fe\,{\sc i} & 4961.92 & 3.63 & -2.25  & 26.5   & 7.42 & 16.4   & 7.03 & 845  & Fe\,{\sc i} & 6301.51 & 3.65 & -0.72  & 112.7  & 7.57 & 106.2  & 7.26 & 816  \\
Fe\,{\sc i} & 4962.58 & 4.18 & -1.18  & 54.0   & 7.51 & 40.3   & 7.09 & 66   & Fe\,{\sc i} & 6315.81 & 4.07 & -1.66  & 39.8   & 7.51 & --     & --   & 1014 \\
Fe\,{\sc i} & 4973.10 & 3.96 & -0.92  & 93.8   & 7.72 & 82.4   & 7.34 & 173  & Fe\,{\sc i} & 6322.69 & 2.59 & -2.43  & 78.6   & 7.69 & 61     & 7.16 & 207  \\
Fe\,{\sc i} & 5002.80 & 3.40 & -1.53  & 78.8   & 7.54 & 81.2   & 7.35 & 687  & Fe\,{\sc i} & 6335.34 & 2.20 & -2.18  & 96.3   & 7.43 & 88.6   & 7.08 & 62   \\
Fe\,{\sc i} & 5022.24 & 3.98 & -0.56  & 99.5   & 7.45 & 93.9   & 7.14 & 965  & Fe\,{\sc i} & 6336.83 & 3.69 & -0.86  & 102.4  & 7.34 & 94.8   & 7.01 & 816  \\
Fe\,{\sc i} & 5029.62 & 3.41 & -2.00  & 48.7   & 7.54 & 40.0   & 7.21 & 718  & Fe\,{\sc i} & 6344.15 & 2.43 & -2.92  & 59.2   & 7.61 & 56.4   & 7.40 & 169  \\
Fe\,{\sc i} & 5074.75 & 4.22 & -0.23  & 118.8  & 7.50 & 105.8  & 7.12 & 1094 & Fe\,{\sc i} & 6393.61 & 2.43 & -1.58  & 134.6  & 7.48 & 113.4  & 6.99 & 168  \\
Fe\,{\sc i} & 5083.35 & 0.96 & -2.96  & 109.9  & 7.43 & 101.8  & 7.02 & 16   & Fe\,{\sc i} & 6408.03 & 3.69 & -1.02  & 97.3   & 7.69 & 88.1   & 7.35 & 816  \\
Fe\,{\sc i} & 5088.16 & 4.15 & -1.75  & 34.8   & 7.59 & 23.5   & 7.22 & 1066 & Fe\,{\sc i} & 6419.96 & 4.73 & -0.27  & 86.8   & 7.51 & 71.3   & 7.11 & 1258 \\
Fe\,{\sc i} & 5141.75 & 2.42 & -2.24  & 89.3   & 7.68 & 76.5   & 7.23 & 114  & Fe\,{\sc i} & 6430.86 & 2.18 & -2.01  & 115.3  & 7.51 & 97.9   & 7.03 & 62   \\
Fe\,{\sc i} & 5145.10 & 2.20 & -3.08$^{\rm *}$  & 52.5   & 7.49 & 49.0   & 7.26 & 66   & Fe\,{\sc i} & 6469.19 & 4.83 & -0.81  & 58.6   & 7.68 & 44.5   & 7.31 & 1258 \\
Fe\,{\sc i} & 5198.72 & 2.22 & -2.13  & 94.9   & 7.48 & 91.1   & 7.17 & 66   & Fe\,{\sc i} & 6481.88 & 2.28 & -2.98  & 64.7   & 7.63 & 62.9   & 7.44 & 109  \\
Fe\,{\sc i} & 5217.40 & 3.21 & -1.16  & 121.7  & 7.52 & 93.8   & 6.94 & 553  & Fe\,{\sc i} & 6498.94 & 0.96 & -4.69  & 45.7   & 7.59 & 40.9   & 7.32 & 13   \\
Fe\,{\sc i} & 5228.38 & 4.22 & -1.26  & 60.0   & 7.68 & 53.5   & 7.41 & 1091 & Fe\,{\sc i} & 6518.37 & 2.83 & -2.46$^{\rm \dagger}$  & 56.0   & 7.46  & --     & --   & 342  \\
Fe\,{\sc i} & 5242.50 & 3.63 & -0.97  & 85.7   & 7.50 & 73.2   & 7.08 & 843  & Fe\,{\sc i} & 6593.88 & 2.43 & -2.42  & 83.7   & 7.60  & 75.0   & 7.25 & 168  \\
Fe\,{\sc i} & 5243.78 & 4.26 & -1.12  & 63.0   & 7.64 & 46.6   & 7.18 & 1089 & Fe\,{\sc i} & 6609.12 & 2.56 & -2.69  & 65.6   & 7.62 & 55.1   & 7.25 & 206  \\
Fe\,{\sc i} & 5247.06 & 0.09 & -4.95  & 65.3   & 7.58 & 61.3   & 7.30 & 1    & Fe\,{\sc i} & 6678.00 & 2.69 & -1.42  & 133.5  & 7.54 & 112.1  & 7.05 & 268  \\
Fe\,{\sc i} & 5250.22 & 0.12 & -4.94  & 65.9   & 7.62 & 61.5   & 7.32 & 1    & Fe\,{\sc i} & 6703.58 & 2.76 & -3.06  & 37.2   & 7.55 & 31.7   & 7.3  & 268  \\
Fe\,{\sc i} & 5250.65 & 2.20 & -2.18  & 100.6  & 7.60 & 95.7   & 7.27 & 66   & Fe\,{\sc i} & 6750.16 & 2.42 & -2.62  & 73.3   & 7.56 & 61.8   & 7.16 & 111  \\
Fe\,{\sc i} & 5253.47 & 3.28 & -1.57  & 77.9   & 7.42 & 62.2   & 6.94 & 553  & Fe\,{\sc ii}& 4178.86 & 2.58 & -2.51$^{\rm *}$  & 89.7   & 7.50 & 77.2   & 7.19 & 28   \\
Fe\,{\sc i} & 5288.53 & 3.69 & -1.51  & 58.6   & 7.52 & 48.4   & 7.15 & 929  & Fe\,{\sc ii}& 4508.29 & 2.85 & -2.44$^{\rm *}$  & 82.1   & 7.46 & 76.6   & 7.30 & 38   \\
Fe\,{\sc i} & 5298.78 & 3.64 & -2.02  & 42.2   & 7.57 & 30.3   & 7.18 & 875  & Fe\,{\sc ii}& 4576.34 & 2.84 & -2.92  & 65.3   & 7.52 & 54.8   & 7.25 & 38   \\
Fe\,{\sc i} & 5307.37 & 1.61 & -2.99  & 87.6   & 7.63 & 80.3   & 7.27 & 36   & Fe\,{\sc ii}& 4582.83 & 2.84 & -3.06  & 58.6   & 7.48 & 42.8   & 7.06 & 37   \\
Fe\,{\sc i} & 5322.05 & 2.28 & -2.80  & 59.4   & 7.44 & 48.9   & 7.04 & 112  & Fe\,{\sc ii}& 4620.52 & 2.83 & -3.19  & 55.3   & 7.50  & 39.8   & 7.10  & 38   \\
Fe\,{\sc i} & 5365.41 & 3.57 & -1.22$^{\rm *}$  & 74.4   & 7.45 & 66.3   & 7.12 & 786  & Fe\,{\sc ii}& 5132.67 & 2.81 & -4.09  & 25.1   & 7.53 & 13.6   & 7.17 & 35   \\
Fe\,{\sc i} & 5373.71 & 4.47 & -0.84  & 61.1   & 7.47 & 49.5   & 7.12 & 1166 & Fe\,{\sc ii}& 5197.58 & 3.23 & -2.22$^{\rm *}$  & 81.6   & 7.51 & 68.3   & 7.20 & 49   \\
Fe\,{\sc i} & 5379.58 & 3.69 & -1.51  & 59.7   & 7.53 & 46.1   & 7.09 & 928  & Fe\,{\sc ii}& 5234.63 & 3.22 & -2.21  & 83.8   & 7.53 & 68.2   & 7.18 & 49   \\
Fe\,{\sc i} & 5398.29 & 4.44 & -0.71  & 71.1   & 7.50 & 60.3   & 7.15 & 553  & Fe\,{\sc ii}& 5264.81 & 3.33 & -3.13$^{\rm *}$  & 45.2   & 7.61 & 29.0   & 7.21 & 48   \\
Fe\,{\sc i} & 5473.91 & 4.15 & -0.79  & 78.2   & 7.49 & 65.1   & 7.08 & 1062 & Fe\,{\sc ii}& 5284.11 & 2.89 & -3.11$^{\rm *}$  & 61.1   & 7.58 & --     & --   & 41   \\
Fe\,{\sc i} & 5483.11 & 4.15 & -1.41  & 46.2   & 7.49 & --     &   -- & 1061 & Fe\,{\sc ii}& 5414.07 & 3.22 & -3.58$^{\rm *}$  & 28.6   & 7.50  & 15.5   & 7.12 & 48   \\
Fe\,{\sc i} & 5487.15 & 4.41 & -1.51  & 36.5   & 7.61 & 26.3   & 7.28 & 1143 & Fe\,{\sc ii}& 5425.26 & 3.20 & -3.22$^{\rm *}$  & 42.5   & 7.50 & 26.2   & 7.09 & 49   \\
Fe\,{\sc i} & 5501.48 & 0.96 & -3.05  & 116.6  & 7.54 & 109.8  & 7.16 & 15   & Fe\,{\sc ii}& 5534.85 & 3.24 & -2.75$^{\rm *}$  & 58.3   & 7.48 & 41.7   & 7.08 & 55   \\
Fe\,{\sc i} & 5506.79 & 0.99 & -2.80  & 121.6  & 7.38 & 115.1  & 7.00 & 15   & Fe\,{\sc ii}& 6247.56 & 3.89 & -2.30$^{\rm *}$  & 54.1   & 7.51 & 40.6   & 7.21 & 74   \\
Fe\,{\sc i} & 5525.55 & 4.23 & -1.08  & 53.3   & 7.37 & 38.1   & 6.94 & 1062 & Fe\,{\sc ii}& 6432.68 & 2.89 & -3.57$^{\rm *}$  & 40.1   & 7.46 & 27.5   & 7.15 & 40   \\
Fe\,{\sc i} & 5543.94 & 4.22 & -1.11  & 62.2   & 7.57 & 48.9   & 7.17 & 1062 & Fe\,{\sc ii}& 6456.39 & 3.90 & -2.05$^{\rm *}$  & 64.6   & 7.50 & 43.9   & 7.04 & 74   \\
Fe\,{\sc i} & 5546.51 & 4.37 & -1.28  & 49.7   & 7.62 & 35.8   & 7.22 & 1145 & Fe\,{\sc ii}& 6516.08 & 2.89 & -3.31$^{\rm *}$  & 54.9   & 7.56 & --     &  --  & 40   \\
Fe\,{\sc i} & 5560.22 & 4.43 & -1.16  & 50.2   & 7.56 & 35.4   & 7.14 & 1164 &  	       &	     &	    &        & 	      &      &        &	     &      \\
 
\hline
\end{tabular}
\end{table*}

\begin{table*}
\setlength{\tabcolsep}{1.5pt}
\small
\caption{Lines used in the analysis of the KPNO Solar spectrum and HD\,218209. Abundances for individual lines are those obtained for a model of $T_{\rm eff}= 5790$ K, $\log g = 4.4$ cgs, and $\xi=$ 0.66 km s$^{\rm −1}$.}
\label{tab:lineslit_other_lines}
\centering
\begin{tabular}{lccccccccccccccccccc}
\hline
 & & & & \multicolumn{2}{c}{Sun} & \multicolumn{2}{c}{HD\,218209}  & & & &  &&   &\multicolumn{2}{c}{Sun} &\multicolumn{2}{c}{HD\,218209}& & \\
\cline{2-4}
\cline{7-8}
\cline{12-14}
\cline{17-18}
 Spec.& $\lambda$ & LEP& $\log(gf)$ & EW & $\log \epsilon$(X) & EW& $\log\epsilon$(X)&RMT&Ref.&Spec.&$\lambda$ & LEP&  $\log(gf)$ & EW     & $\log \epsilon$(X) & EW	& $\log\epsilon$(X) & RMT &Ref.\\
\cline{2-4}
\cline{7-8}
\cline{12-14}
\cline{17-18}
   & (\AA)  & (eV) & (dex)   &(m\AA) & (dex) &(m\AA) & (dex)&    &    && (\AA)  & (eV) & (dex)   &(m\AA) & (dex) &(m\AA) & (dex)&\\
\hline

Na\,{\sc I}  & 5682.65 &  2.10 & -0.67  & 100.1 & 6.21 & 82.4	&  5.86 & 6    & 1 & Ti\,{\sc i}  & 5145.47 & 1.46 & -0.54  & 37.2  & 4.94 & 31.4  & 4.67 & 109  & 7 \\
Na\,{\sc I}  & 5688.22 &  2.10 & -0.37  & 121.1 & 6.11 & 99.9	&  5.72 & 6    & 1 & Ti\,{\sc i}  & 5147.48 & 0.00 & -1.94  & 38.8  & 4.93 & 37.7  & 4.73 & 4	 & 7 \\
Mg\,{\sc i}  & 4571.10 &  0.00 & -5.40  & 109.2 & 7.54 & 115.1  &  7.44 & 1    & 2 & Ti\,{\sc i}  & 5152.19 & 0.02 & -1.95  & 36.9  & 4.92 & --    & --   & 4	 & 7 \\
Mg\,{\sc i}  & 5711.10 &  4.34 & -1.74  & 104.6 & 7.65 & 107.6  &  7.52 & 8    & 2 & Ti\,{\sc i}  & 5192.98 & 0.02 & -0.95  & 84.0  & 5.03 & 85.2  & 4.83 & 4	 & 7 \\
Si\,{\sc i}  & 5645.62 &  4.93 & -2.03  & 35.5  & 7.50 & --	    &   --  & 10   & 3 & Ti\,{\sc i}  & 5210.39 & 0.05 & -0.82  & 89.0  & 5.03 & 98.6  & 4.95 & 4	 & 7 \\
Si\,{\sc i}  & 5665.56 &  4.92 & -1.99  & 39.6  & 7.53 & 31.0	&  7.37 & 10   & 3 & Ti\,{\sc i}  & 5219.71 & 0.02 & -2.22  & 29.1  & 5.00 & 29.0  & 4.82 & 4	 & 7 \\
Si\,{\sc i}  & 5684.49 &  4.95 & -1.58  & 60.6  & 7.52 & 54.3	&  7.35 & 11   & 3 & Ti\,{\sc i}  & 5490.16 & 1.46 & -0.84  & 22.4  & 4.85 & 25.6  & 4.80 & 3	 & 7 \\
Si\,{\sc i}  & 5708.40 &  4.95 & -1.47  & 74.7  & 7.64 & 62.2	&  7.37 & 10   & 4 & Ti\,{\sc i}  & 5866.46 & 1.07 & -0.79  & 46.9  & 4.97 & 44.9  & 4.78 & 72   & 7 \\
Si\,{\sc i}  & 5772.15 &  5.08 & -1.62  & 52.0  & 7.53 & 40.0	&  7.27 & 17   & 3 & Ti\,{\sc i}  & 6126.22 & 1.07 & -1.42  & 21.7  & 5.00 & 23.0  & 4.88 & 69   & 7 \\
Si\,{\sc i}  & 5793.08 &  4.93 & -1.86  & 42.4  & 7.46 & 31.6	&  7.20 & 9    & 3 & Ti\,{\sc i}  & 6258.11 & 1.44 & -0.39  & 51.4  & 5.00 & 48.3  & 4.78 & 104  & 7 \\
Si\,{\sc i}  & 5948.54 &  5.08 & -1.09  & 83.8  & 7.51 & --   	&  --	& 16   & 3 & Ti\,{\sc i}  & 6261.11 & 1.43 & -0.53  & 49.9  & 5.10 & 53.2  & 5.02 & 104  & 7 \\
Si\,{\sc i}  & 6125.03 &  5.61 & -1.53  & 30.9  & 7.51 & 23.1	&  7.31 & 30   & 3 & Ti\,{\sc i}  & 6336.11 & 1.44 & -1.69  & 5.60  & 4.92 & --    & --   & 103  & 7 \\
Si\,{\sc i}  & 6142.49 &  5.62 & -1.48  & 33.3  & 7.51 & 26.8	&  7.35 & 30   & 3 & Ti\,{\sc i}  & 6743.13 & 0.90 & -1.63  & 18.0  & 4.89 & --    & --   & 48   & 7 \\
Si\,{\sc i}  & 6145.02 &  5.61 & -1.39  & 37.9  & 7.50 & 27.6	&  7.26 & 29   & 3 & Ti\,{\sc ii} & 4443.81 & 1.08 & -0.71  & 139.7 & 5.04 & 139.9 & 4.86 & 19   & 7 \\
Si\,{\sc i}  & 6244.48 &  5.61 & -1.29  & 44.0  & 7.51 & 32.9	&  7.27 & 27   & 3 & Ti\,{\sc ii} & 4468.50 & 1.13 & -0.63  & 134.1 & 4.94 & 149.8 & 4.90 & 31   & 7 \\
Si\,{\sc i}  & 6721.84 &  5.86 & -0.94  & 42.3  & 7.33 & 29.1	&  7.06 & 38*  & 4 & Ti\,{\sc ii} & 4493.53 & 1.08 & -2.78  & 34.2  & 4.90 & 40.1  & 5.01 & 18   & 7 \\
Ca\,{\sc i}  & 4512.27 &  2.52 & -1.90  & 25.0  & 6.33 & 16.3	&  5.99 & 24   & 5 & Ti\,{\sc ii} & 4568.33 & 1.22 & -2.65  & 32.0  & 4.84 & 26.7  & 4.68 & 60   & 7 \\
Ca\,{\sc i}  & 4578.56 &  2.52 & -0.70  & 85.8  & 6.33 & 89.4	&  6.18 & 23   & 5 & Ti\,{\sc ii} & 4583.41 & 1.16 & -2.84  & 31.7  & 4.96 & 25.6  & 4.77 & 39   & 7 \\
Ca\,{\sc i}  & 5260.39 &  2.52 & -1.72  & 33.2  & 6.32 & 25.2	&  6.04 & 22   & 5 & Ti\,{\sc ii} & 4708.67 & 1.24 & -2.35  & 51.0  & 5.02 & 44.0  & 4.82 & 49   & 7 \\
Ca\,{\sc i}  & 5261.71 &  2.52 & -0.58  & 98.7  & 6.49 & 91.7	&  6.19 & 22   & 5 & Ti\,{\sc ii} & 4874.01 & 3.09 & -0.86  & 36.4  & 4.91 & 27.3  & 4.70 & 114  & 7 \\
Ca\,{\sc i}  & 5512.99 &  2.93 & -0.46  & 83.9  & 6.36 & 82.2	&  6.16 & 48   & 5 & Ti\,{\sc ii} & 4911.20 & 3.12 & -0.64  & 51.4  & 5.11 & 48.7  & 5.06 & 114  & 7 \\  
Ca\,{\sc i}  & 5581.98 &  2.52 & -0.56  & 94.3  & 6.38 & 88.8	&  6.11 & 21   & 5 & Ti\,{\sc ii} & 5005.17 & 1.57 & -2.73  & 23.9  & 5.02 & 16.2  & 4.76 & 71   & 7 \\  
Ca\,{\sc i}  & 5590.13 &  2.52 & -0.57  & 93.0  & 6.37 & 86.6	&  6.09 & 21   & 5 & Ti\,{\sc ii} & 5013.69 & 1.58 & -2.14  & 49.5  & 5.08 & 43.7  & 4.92 & 71   & 7 \\  
Ca\,{\sc i}  & 5601.29 &  2.52 & -0.52  & 102.8 & 6.47 & 99.0	&  6.21 & 21   & 5 & Ti\,{\sc ii} & 5336.79 & 1.58 & -1.60  & 72.3  & 5.09 & 71.0  & 5.03 & 69   & 7 \\  
Ca\,{\sc i}  & 6102.73 &  1.88 & -0.81  & 122.3 & 6.10 & 123.8  &  5.86 & 3    & 5 & Ti\,{\sc ii} & 5418.77 & 1.58 & -2.13  & 48.7  & 5.02 & 41.3  & 4.83 & 69   & 7 \\  
Ca\,{\sc i}  & 6166.44 &  2.52 & -1.14  & 72.3  & 6.37 & 64.2	&  6.08 & 20   & 5 & V\,{\sc i}   & 4437.84 & 0.29 & -0.71  & 37.1  & 4.03 & 32.9  & 3.76 & 21   & 8 \\  
Ca\,{\sc i}  & 6169.04 &  2.52 & -0.80  & 93.5  & 6.34 & 90.8	&  6.09 & 20   & 5 & V\,{\sc i}   & 4577.18 & 0.00 & -1.08  & 33.6  & 4.01 & 36.5  & 3.91 & 4	 & 8 \\  
Ca\,{\sc i}  & 6169.56 &  2.52 & -0.48  & 114.0 & 6.26 & 109.4  &  5.98 & 20   & 5 & V\,{\sc i}   & 5727.06 & 1.08 & -0.02  & 38.7  & 4.04 & --    & --   & 35   & 8 \\  
Ca\,{\sc i}  & 6455.60 &  2.52 & -1.34  & 56.7  & 6.39 & 47.3	&  6.09 & 19   & 5 & V\,{\sc i}   & 6119.53 & 1.06 & -0.36  & 22.1  & 3.93 & 21.8  & 3.78 & 34   & 8 \\  
Ca\,{\sc i}  & 6471.67 &  2.52 & -0.69  & 89.3  & 6.35 & 85.2	&  6.12 & 18   & 5 & V\,{\sc i}   & 6243.11 & 0.30 & -0.94  & 29.6  & 3.94 & 29.9  & 3.78 & 19   & 8 \\  
Ca\,{\sc i}  & 6493.79 &  2.52 & -0.11  & 124.4 & 6.26 & 116.6  &  5.95 & 18   & 5 & Cr\,{\sc i}  & 4545.96 & 0.94 & -1.38  & 81.6  & 5.68 & 80.4  & 5.43 & 10   & 4 \\  
Ca\,{\sc i}  & 6499.65 &  2.52 & -0.82  & 82.5  & 6.37 & 81.2	&  6.19 & 18   & 5 & Cr\,{\sc i}  & 4616.13 & 0.98 & -1.18  & 90.3  & 5.71 & 83.3  & 5.34 & 21   & 4 \\  
Ca\,{\sc i}  & 6572.79 &  0.00 & -4.32  & 32.1  & 6.37 & 33.4	&  6.24 & 1    & 5 & Cr\,{\sc i}  & 4626.18 & 0.97 & -1.32  & 82.2  & 5.66 & 72.7  & 5.24 & 21   & 4 \\  
Ca\,{\sc i}  & 6717.69 &  2.71 & -0.52  & 99.5  & 6.36 & 98.7	&  6.04 & 32   & 5 & Cr\,{\sc i}  & 4646.17 & 1.03 & -0.71  &105.5  & 5.55 & 94.3  & 5.11 & 21   & 4 \\  
Sc\,{\sc i}  & 4023.69 &  0.02 & 0.38	& 57.5  & 3.12 & 51.3	&  2.78 & 7    & 6 & Cr\,{\sc i}  & 4651.29 & 0.98 & -1.46  & 82.9  & 5.82 & 66.1  & 5.23 & 21   & 4 \\  
Sc\,{\sc ii} & 4246.84 &  0.31 & 0.24	& 156.6 & 3.17 & 153.8  &  2.94 & 7    & 6 & Cr\,{\sc i}  & 4652.17 & 1.00 & -1.03  &100.8  & 5.77 & 86.3  & 5.26 & 21   & 4 \\  
Sc\,{\sc ii} & 5239.82 &  1.45 & -0.76  & 52.0  & 3.22 & 48.7	&  3.12 & 26   & 6 & Cr\,{\sc i}  & 4708.02 & 3.17 & 0.11   & 57.5  & 5.58 & 41.5  & 5.11 & 186  & 4 \\  
Sc\,{\sc ii} & 5526.82 &  1.77 & -0.01  & 75.1  & 3.32 & 65.4	&  3.06 & 31   & 6 & Cr\,{\sc i}  & 4718.42 & 3.19 & 0.10   & 64.1  & 5.74 & 55.9  & 5.42 & 186  & 4 \\  
Sc\,{\sc ii} & 5640.99 &  1.50 & -0.99  & 39.7  & 3.18 & 34.7	&  3.04 & 29   & 6 & Cr\,{\sc i}  & 4730.72 & 3.08 & -0.19  & 48.3  & 5.67 & 30.5  & 5.12 & 145  & 4 \\  
Sc\,{\sc ii} & 5657.88 &  1.51 & -0.54  & 66.1  & 3.36 & 55.6	&  3.09 & 29   & 6 & Cr\,{\sc i}  & 4737.35 & 3.07 & -0.10  & 55.7  & 5.68 & 43.0  & 5.26 & 145  & 4 \\  
Sc\,{\sc ii} & 5667.15 &  1.50 & -1.21  & 32.1  & 3.22 & 26.1	&  3.05 & 29   & 6 & Cr\,{\sc i}  & 4756.12 & 3.10 & 0.09   & 62.9  & 5.76 & 52.3  & 5.39 & 145  & 4 \\  
Sc\,{\sc ii} & 5669.04 &  1.50 & -1.10  & 34.1  & 3.16 & 29.7	&  3.03 & 29   & 6 & Cr\,{\sc i}  & 4936.34 & 3.11 & -0.34  & 46.2  & 5.77 & 37.6  & 5.46 & 166  & 4 \\  
Ti\,{\sc i}  & 4060.27 &  1.05 & -0.69  & 37.8  & 4.82 & 44.5	&  4.83 & 80   & 7 & Cr\,{\sc i}  & 4964.93 & 0.94 & -2.53  & 37.7  & 5.65 & 33.3  & 5.40 & 9	 & 4 \\  
Ti\,{\sc i}  & 4186.13 &  1.50 & -0.24  & 42.7  & 4.91 & 43.7	&  4.79 & 129  & 7 & Cr\,{\sc i}  & 5247.57 & 0.96 & -1.63  & 82.0  & 5.78 & 74.5  & 5.42 & 18   & 4 \\  
Ti\,{\sc i}  & 4287.41 &  0.84 & -0.37  & 75.3  & 5.20 & 76.7	&  5.01 & 44   & 7 & Cr\,{\sc i}  & 5296.70 & 0.98 & -1.41  & 91.9  & 5.77 & 80.7  & 5.33 & 18   & 4 \\  
Ti\,{\sc i}  & 4453.32 &  1.43 & -0.03  & 64.5  & 5.16 &  --	&  --	& 113  & 7 & Cr\,{\sc i}  & 5300.75 & 0.98 & -2.13  & 58.6  & 5.75 & 44.0  & 5.25 & 18   & 4 \\ 
Ti\,{\sc i}  & 4465.81 &  1.74 &  -0.13 & 40.9  & 4.95 & 40.7	& 4.81  & 146  & 7 & Cr\,{\sc i}  & 5345.81 & 1.00 & -0.98  &116.8  & 5.73 & 111.0 & 5.37 & 18   & 4 \\ 
Ti\,{\sc i}  & 4512.74 &  0.84 &  -0.40 & 67.5  & 4.99 & 68.4	& 4.82  & 42   & 7 & Cr\,{\sc i}  & 5348.33 & 1.00 & -1.29  & 99.5  & 5.79 & 86.5  & 5.32 & 18   & 4 \\ 
Ti\,{\sc i}  & 4518.03 &  0.83 &  -0.25 & 73.0  & 4.97 & 74.6	& 4.80  & 42   & 7 & Cr\,{\sc i}  & 5787.93 & 3.32 & -0.08  & 45.7  & 5.62 & 32.3  & 5.22 & 188  & 4 \\ 
Ti\,{\sc i}  & 4534.79 &  0.84 &  0.35  & 95.9  & 4.86 & 96.5	& 4.60  & 42   & 7 & Cr\,{\sc ii} & 4588.20 & 4.07 & -0.65  & 71.8  & 5.69 & 59.4  & 5.40 & 44   & 9 \\ 
Ti\,{\sc i}  & 4548.77 &  0.83 &  -0.28 & 70.9  & 4.95 & 82.8	& 4.99  & 42   & 7 & Cr\,{\sc ii} & 5237.32 & 4.07 & -1.17  & 53.0  & 5.75 & 39.2  & 5.42 & 43   & 9 \\ 
Ti\,{\sc i}  & 4555.49 &  0.85 &  -0.40 & 63.1  & 4.89 & 64.5	& 4.74  & 42   & 7 & Cr\,{\sc ii} & 5305.87 & 3.83 & -1.91  & 24.6  & 5.48 & 13.2  & 5.13 & 24   & 9 \\ 
Ti\,{\sc i}  & 4617.28 &  1.75 &  0.44  & 63.7  & 4.94 & 61.3	& 4.72  & 145  & 7 & Mn\,{\sc i}  & 4055.55 & 2.14 & -0.08  &125.5  & 5.76 & 107.2 & 5.26 & 5	 & 10\\ 
Ti\,{\sc i}  & 4623.10 &  1.74 &  0.16  & 56.6  & 5.03 & 52.7	& 4.79  & 145  & 7 & Mn\,{\sc i}  & 4082.94 & 2.18 & -0.36  & 91.3  & 5.60 & 76.6  & 5.07 & 5	 & 10\\ 
Ti\,{\sc i}  & 4639.36 &  1.74 &  -0.05 & 43.2  & 4.90 & 49.4	& 4.91  & 145  & 7 & Mn\,{\sc i}  & 4451.59 & 2.89 & 0.28   & 95.1  & 5.55 & 80.0  & 5.06 & 22   & 10\\ 
Ti\,{\sc i}  & 4639.66 &  1.75 &  -0.14 & 43.5  & 5.01 & 44.4	& 4.89  & 145  & 7 & Mn\,{\sc i}  & 4470.14 & 2.94 & -0.44  & 56.9  & 5.46 & 36.7  & 4.82 & 22   & 10\\ 
Ti\,{\sc i}  & 4656.47 &  0.00 &  -1.28 & 68.8  & 5.07 & 76.2	& 5.05  & 6    & 7 & Mn\,{\sc i}  & 4502.22 & 2.92 & -0.34  & 59.6  & 5.41 & 43.1  & 4.86 & 22   & 10\\ 
Ti\,{\sc i}  & 4722.61 &  1.05 &  -1.47 & 18.3  & 5.02 & 18.7	& 4.88  & 75   & 7 & Mn\,{\sc i}  & 4709.72 & 2.89 & -0.49  & 68.4  & 5.73 & 44.9  & 5.01 & 21   & 10\\ 
Ti\,{\sc i}  & 4742.80 &  2.24 &  0.21  & 31.6  & 4.85 & 32.5	& 4.74  & 233  & 7 & Mn\,{\sc i}  & 4739.11 & 2.94 & -0.61  & 59.3  & 5.66 & 38.7  & 5.01 & 21   & 10\\ 
Ti\,{\sc i}  & 4758.12 &  2.25 &  0.51  & 44.9  & 4.87 & 39.8	& 4.63  & 233  & 7 & Mn\,{\sc i}  & 4765.86 & 2.94 & -0.09  & 77.4  & 5.57 & 66.0  & 5.15 & 21   & 10\\ 
Ti\,{\sc i}  & 4759.28 &  2.25 &  0.59  & 44.8  & 4.79 & 43.9	& 4.64  & 233  & 7 & Mn\,{\sc i}  & 4766.42 & 2.92 & 0.10   & 93.8  & 5.68 & 76.3  & 5.16 & 21   & 10\\ 
Ti\,{\sc i}  & 4820.41 &  1.50 &  -0.38 & 42.3  & 4.96 & 39.9	& 4.77  & 126  & 7 & Mn\,{\sc i}  & 4783.42 & 2.30 & 0.03   &140.1  & 5.75 & 121.1 & 5.27 & 16   & 10\\ 
Ti\,{\sc i}  & 4885.09 &  1.89 &  0.41  & 62.6  & 5.04 & 70.3	& 5.05  & 231  & 7 & Mn\,{\sc i}  & 5117.94 & 3.13 & -1.20  & 24.0  & 5.51 & --    & --   & 32   & 10\\ 
Ti\,{\sc i}  & 4913.62 &  1.87 &  0.22  & 50.6  & 4.91 & 48.1	& 4.71  & 157  & 7 & Mn\,{\sc i}  & 5432.55 & 0.00 & -3.79  & 49.6  & 5.61 & 25.9  & 4.86 & 1	 & 10\\ 
Ti\,{\sc i}  & 4981.74 &  0.85 &  0.57  & 119.3 & 4.92 & 120.6  & 4.64  & 38   & 7 & Mn\,{\sc i}  & 6021.80 & 3.07 & -0.05  & 96.9  & 5.85 & 69.4  & 5.17 & 27   & 10\\ 
Ti\,{\sc i}  & 4999.51 &  0.83 &  0.32  & 102.5 & 4.91 & 104.8  & 4.67  & 38   & 7 & Co\,{\sc i}  & 4121.33 & 0.92 & -0.33  &124.0  & 5.06 & 113.1 & 4.61 & 28   & 11\\ 
Ti\,{\sc i}  & 5009.65 &  0.02 &  -2.20 & 24.2  & 4.88 & 34.5	& 4.95  & 5    & 7 & Co\,{\sc i}  & 4813.48 & 3.21 & 0.12   & 45.4  & 5.02 &  31.1 & 4.56 & 158  & 11\\
Ti\,{\sc i}  & 5016.17 &  0.85 &  -0.48 & 64.8  & 4.94 & 64.6	& 4.76  & 38   & 7 & Co\,{\sc i}  & 5301.05 & 1.71 & -1.90  & 20.6  & 4.90 &  --   & --   & 39   & 11\\
Ti\,{\sc i}  & 5020.03 &  0.83 &  -0.33 & 77.6  & 5.07 & 79.9	& 4.91  & 38   & 7 & Co\,{\sc i}  & 5342.71 & 4.02 & 0.74   & 30.5  & 4.77 &  21.2 & 4.43 & 190  & 11\\
Ti\,{\sc i}  & 5022.87 &  0.83 &  -0.33 & 73.2  & 4.97 & 73.6	& 4.78  & 38   & 7 & Co\,{\sc i}  & 5352.05 & 3.58 & 0.06   & 24.2  & 4.85 &  16.6 & 4.52 & 172  & 11\\
Ti\,{\sc i}  & 5039.96 &  0.02 &  -1.08 & 73.8  & 4.95 & 79.9	& 4.87  & 5    & 7 & Co\,{\sc i}  & 5483.36 & 1.71 & -1.50  & 49.8  & 5.23 &  39.6 & 4.84 & 39   & 11\\
Ti\,{\sc i}  & 5064.65 &  0.05 &  -0.94 & 84.8  & 5.09 & 87.7	& 4.92  & 5    & 7 & Co\,{\sc i}  & 6093.15 & 1.74 & -2.40  &  9.2  & 4.95 &  8.1  & 4.75 & 37   & 11\\
\hline
\end{tabular}
\end{table*}

\begin{table*}
\setcounter{table}{2}
\centering
{\bf Table 2 continued}\\
\small
\setlength{\tabcolsep}{1.5pt}
\label{tab:lineslit_other_lines_3}
\begin{tabular}{lccccccccccccccccccc}
\hline
 & & & & \multicolumn{2}{c}{Sun} & \multicolumn{2}{c}{HD\,218209}  & & & &  &&   &\multicolumn{2}{c}{Sun} &\multicolumn{2}{c}{HD\,218209}&  & \\
\cline{2-4}
\cline{7-8}
\cline{12-14}
\cline{17-18}
 Spec.& $\lambda$ & LEP& $\log(gf)$ & EW & $\log \epsilon$(X) & EW& $\log\epsilon$(X)&RMT&Ref.&Spec.&$\lambda$ & LEP&  $\log(gf)$ & EW     & $\log \epsilon$(X) & EW	& $\log\epsilon$(X) & RMT &Ref.\\
\cline{2-4}
\cline{7-8}
\cline{12-14}
\cline{17-18}
   & (\AA)  & (eV) & (dex)   &(m\AA) & (dex) &(m\AA) & (dex)&    &    && (\AA)  & (eV) & (dex)   &(m\AA) & (dex) &(m\AA) & (dex)&\\
\hline
Ni\,{\sc i} & 4410.52 & 3.31 & -1.08  & 57.8  & 6.39 &  43.1  & 5.94 & 88  & 4 & Ni\,{\sc i} & 5748.36 & 1.68 & -3.26  & 28.8  & 6.26 &  19.3  & 5.87 & 45  & 4 \\
Ni\,{\sc i} & 4470.48 & 3.40 & -0.40  & 79.5  & 6.24 &  76.2  & 5.98 & 86  & 4 & Ni\,{\sc i} & 5805.23 & 4.17 & -0.64  & 40.5  & 6.30 &  28.2  & 5.93 & 234 & 4 \\
Ni\,{\sc i} & 4606.23 & 3.60 & -1.02  & 48.0  & 6.37 &  38.8  & 6.05 & 100 & 4 & Ni\,{\sc i} & 6007.32 & 1.68 & -3.34  & 25.4  & 6.24 &  22.5  & 6.04 & 42  & 4 \\
Ni\,{\sc i} & 4686.22 & 3.60 & -0.64  & 62.3  & 6.30 &  58.9  & 6.08 & 98  & 4 & Ni\,{\sc i} & 6086.29 & 4.26 & -0.51  & 43.5  & 6.30 &  28.2  & 5.88 & 249 & 4 \\
Ni\,{\sc i} & 4731.80 & 3.83 & -0.85  & 45.4  & 6.35 &  26.6  & 5.82 & 163 & 4 & Ni\,{\sc i} & 6108.12 & 1.68 & -2.44  & 65.6  & 6.30 &  55.3  & 5.92 & 45  & 4 \\
Ni\,{\sc i} & 4732.47 & 4.10 & -0.55  & 44.3  & 6.29 &  30.6  & 5.89 & 235 & 4 & Ni\,{\sc i} & 6128.98 & 1.68 & -3.32  &25.3   & 6.21 & 20.1   & 5.94 & 42  & 4 \\ 
Ni\,{\sc i} & 4752.43 & 3.66 & -0.69  & 59.1  & 6.33 &  43.1  & 5.86 & 132 & 4 & Ni\,{\sc i} & 6130.14 & 4.26 & -0.96  &21.6   & 6.24 & 11.5   & 5.82 & 248 & 4 \\ 
Ni\,{\sc i} & 4756.52 & 3.48 & -0.34  & 77.5  & 6.19 &  76.1  & 5.96 & 98  & 4 & Ni\,{\sc i} & 6175.37 & 4.09 & -0.54  &50.7   & 6.32 & 36.4   & 5.92 & 217 & 4 \\ 
Ni\,{\sc i} & 4806.99 & 3.68 & -0.64  & 61.4  & 6.35 &  54.7  & 6.07 & 163 & 4 & Ni\,{\sc i} & 6176.82 & 4.09 & -0.53  &63.0   & 6.54 & 48.0   & 6.13 & 228 & 4 \\ 
Ni\,{\sc i} & 4829.03 & 3.54 & -0.33  & 80.1  & 6.24 &  68.7  & 5.85 & 131 & 4 & Ni\,{\sc i} & 6204.61 & 4.09 & -1.14  &21.7   & 6.26 & 18.2   & 6.08 & 226 & 4 \\ 
Ni\,{\sc i} & 4852.56 & 3.54 & -1.07  & 44.7  & 6.28 &  42.3  & 6.10 & 130 & 4 & Ni\,{\sc i} & 6322.17 & 4.15 & -1.17  &18.3   & 6.24 & --     & --   & 249 & 4 \\ 
Ni\,{\sc i} & 4904.42 & 3.54 & -0.17  & 87.1  & 6.19 &  78.1  & 5.84 & 129 & 4 & Ni\,{\sc i} & 6327.60 & 1.68 & -3.15  &37.9   & 6.34 & 27.8   & 5.97 & 44  & 4 \\ 
Ni\,{\sc i} & 4913.98 & 3.74 & -0.62  & 55.3  & 6.24 &  42.3  & 5.84 & 132 & 4 & Ni\,{\sc i} & 6378.26 & 4.15 & -0.90  &31.5   & 6.32 & 23.5   & 6.04 & 247 & 4 \\ 
Ni\,{\sc i} & 4935.83 & 3.94 & -0.36  & 62.9  & 6.30 &  48.5  & 5.88 & 177 & 4 & Ni\,{\sc i} & 6414.59 & 4.15 & -1.21  &16.8   & 6.23 & --     & --   & 244 & 4 \\ 
Ni\,{\sc i} & 4946.03 & 3.80 & -1.29  & 28.0  & 6.35 &  17.0  & 5.96 & 148 & 4 & Ni\,{\sc i} & 6482.81 & 1.93 & -2.63  &40.8   & 6.12 & 31.8   & 5.79 & 66  & 4 \\ 
Ni\,{\sc i} & 4953.21 & 3.74 & -0.66  & 56.2  & 6.30 &  40.9  & 5.85 & 111 & 4 & Ni\,{\sc i} & 6598.61 & 4.23 & -0.98  &24.7   & 6.30 & --     & --   & 249 & 4 \\ 
Ni\,{\sc i} & 4998.23 & 3.61 & -0.78  & 57.5  & 6.33 &  43.2  & 5.90 & 111 & 4 & Ni\,{\sc i} & 6635.14 & 4.42 & -0.83  &24.0   & 6.31 & 15.2   & 5.97 & 264 & 4 \\ 
Ni\,{\sc i} & 5010.94 & 3.63 & -0.87  & 48.4  & 6.23 &  33.4  & 5.79 & 144 & 4 & Ni\,{\sc i} & 6767.78 & 1.83 & -2.17  &79.6   & 6.47 & 67.4   & 6.06 & 57  & 4 \\ 
Ni\,{\sc i} & 5032.73 & 3.90 & -1.27  & 23.8  & 6.31 &  16.2  & 6.00 & 207 & 4 & Ni\,{\sc i} & 6772.32 & 3.66 & -0.99  &48.4   & 6.27 & 35.3   & 5.89 & 127 & 4 \\ 
Ni\,{\sc i} & 5035.37 & 3.63 &  0.29  & 107.0 & 6.05 &  91.5  & 5.63 & 143 & 4 & Zn\,{\sc i} & 4722.16 & 4.03 & -0.39  &66.1   & 4.70 & 65.0   & 4.58 & 2   & 12\\ 
Ni\,{\sc i} & 5042.19 & 3.64 & -0.57  & 61.8  & 6.21 &  52.6  & 5.88 & 131 & 4 & Zn\,{\sc i} & 4810.54 & 4.08 &  -0.17 & 72.3  & 4.66 &  73.6  & 4.58 & 2   & 12\\
Ni\,{\sc i} & 5048.85 & 3.85 & -0.37  & 66.1  & 6.29 &  51.5  & 5.86 & 195 & 4 & Sr\,{\sc i} & 4607.34 & 0.00 &  0.28  & 45.4  & 2.91 &  32.0  & 2.28 & 2   & 13\\
Ni\,{\sc i} & 5082.35 & 3.66 & -0.54  & 62.5  & 6.22 &  55.2  & 5.92 & 130 & 4 & Y\,{\sc ii} & 4883.69 & 1.08 &  0.07  & 57.6  & 2.33 &  53.4  & 2.15 & 22  & 14\\
Ni\,{\sc i} & 5084.10 & 3.68 &  0.03  & 91.1  & 6.14 &  78.2  & 5.75 & 162 & 4 & Y\,{\sc ii} & 5087.43 & 1.08 &  -0.17 & 48.3  & 2.25 &  31.1  & 1.69 & 20  & 14\\
Ni\,{\sc i} & 5088.54 & 3.85 & -0.91  & 31.9  & 6.12 &  23.9  & 5.83 & 190 & 4 & Zr\,{\sc ii}& 4208.98 & 0.71 &  -0.46 & 44.7  & 2.68 &  39.3  & 2.45 & 41  & 15\\
Ni\,{\sc i} & 5102.97 & 1.68 & -2.62  & 48.1  & 6.16 &  36.7  & 5.74 & 49  & 4 & Ba\,{\sc ii}& 4554.04 & 0.00 &  0.14  & 174.6 & 2.17 &  154.7 & 1.78 & 1   & 16\\
Ni\,{\sc i} & 5115.40 & 3.83 & -0.11  & 75.2  & 6.19 &  68.9  & 5.90 & 177 & 4 & Ba\,{\sc ii}& 5853.69 & 0.60 &  -0.91 & 62.8  & 2.28 &  51.2  & 1.87 & 2   & 16\\
Ni\,{\sc i} & 5155.13 & 3.90 & -0.66  & 49.2  & 6.28 &  34.5  & 5.86 & 206 & 4 & Ba\,{\sc ii}& 6141.71 & 0.70 &  -0.03 & 112.2 & 2.23 &  98.9  & 1.87 & 2   & 16\\
Ni\,{\sc i} & 5435.87 & 1.99 & -2.60  & 52.0  & 6.52 &  40.0  & 6.09 & 70  & 4 & Ba\,{\sc ii}& 6496.91 & 0.60 &  -0.41 & 97.5  & 2.30 &  90.6  & 2.01 & 2   & 16\\
Ni\,{\sc i} & 5587.87 & 1.93 & -2.14  & 55.8  & 6.08 &  39.7  & 5.56 & 70  & 4 & Ce\,{\sc ii}& 4562.37 & 0.48 &  0.21  & 21.9  & 1.65 &  18.6  & 1.47 & 1   & 17\\
Ni\,{\sc i} & 5593.75 & 3.90 & -0.84  & 45.4  & 6.35 &  28.3  & 5.88 & 206 & 4 & Ce\,{\sc ii}& 4628.16 & 0.52 &  0.14  & 18.4  & 1.63 &  12.9  & 1.34 & 1   & 17\\
Ni\,{\sc i} & 5625.33 & 4.09 & -0.70  & 38.1  & 6.24 &  28.8  & 5.94 & 221 & 4 & Nd\,{\sc ii}& 4446.40 & 0.20 &  -0.35 & 12.7  & 1.41 &   8.0   &  1.08  & 49  & 18\\
Ni\,{\sc i} & 5637.12 & 4.09 & -0.80  & 33.3  & 6.23 &  --    & --   & 218 & 4 & Nd\,{\sc ii}& 5092.80 & 0.38 &  -0.61 & 7.0   & 1.48 &   --   &  --  & 48  & 18\\
Ni\,{\sc i} & 5641.89 & 4.10 & -1.08  & 24.6  & 6.31 &  13.1  & 5.87 & 234 & 4 & Nd\,{\sc ii}& 5293.17 & 0.82 &  0.10  & 10.3  & 1.39 &   --   &  --  & 75  & 18\\
Ni\,{\sc i} & 5682.21 & 4.10 & -0.47  & 51.9  & 6.30 &  33.9  & 5.83 & 232 & 4 & Sm\,{\sc ii}& 4577.69 & 0.25 &  -0.65 & 5.2   & 0.96 &   --   &  --  & 23  & 19\\

\hline
\end{tabular}

\begin{list}{}{}
\item References for the adopted $gf$-values: (1) \citet{Takeda2003}, (2) \citet{Rhodin2017}, (3) \citet{shi2011}, (4) NIST Atomic Spectra Database {(http://physics.nist.gov/PhysRefData/ASD)}, (5) \citet{DenHartog2021}, (6) \citet{Lawler2019}, (7) \citet{Lawler2013}, (8) \citet{Lawler2014}, (9) \citet{Lawler2017}, (10) \citet{DenHartog2011}, (11) \citet{Lawler2015}, (12) \citet{Biemont1980}, (13) \citet{Hansen2013}, (14) \citet{Hannaford1982}, (15) \citet{Biemont1981}, (16) \citet{Klose2002}, (17) \citet{Lawler2009}, (18)  \citet{DenHartog2003}, (19) \citet{Lawler2006}, ($\dagger$) \citet{Bard1994}, (*) \citet{Melendez2009}
\end{list}
\end{table*} 

\begin{table*}
\setlength{\tabcolsep}{5pt}
\renewcommand{\arraystretch}{1.1}
\small
\caption{Solar abundances obtained by employing the Solar model atmosphere from \citet{Castelli2003} compared to
the photospheric abundances from \citet{asplund2009}. The abundances were calculated using the line EWs.}
\label{tab:solar_abund}
\centering
\begin{tabular}{lcccccccc}
\hline
     	&  \multicolumn{4}{c}{HD\,218209}   & \multicolumn{2}{c}{Sun$^{\rm \dagger}$}  &   \citet{asplund2009} &  \\
\cline{2-5}
\cline{8-8}
Species   & $\log\epsilon$(X)$^{\rm 1}$ & [X/Fe]$^{\rm 1}$ & [X/Fe]$^{\rm 2}$ & $N^{\rm 1}$ &  $\log\epsilon_{\rm \odot}$(X) & $N$ &   $\log\epsilon_{\rm \odot}$(X) & $\Delta\log\epsilon_{\rm \odot}$(X) \\
\cline{2-5}
\cline{8-8}
		&  (dex) & (dex) &  (dex) &  & &  & (dex) & \\
 \hline
Na\,{\sc i}  & 5.79$\pm$0.09 &-0.02$\pm$0.14&  0.00$\pm$0.18 & 2   &  6.16$\pm$0.07  &    2	 &   6.24$\pm$0.04  & 0.08  \\
Mg\,{\sc i}  & 7.48$\pm$0.06 & 0.23$\pm$0.13&  0.09$\pm$0.20 & 2   &  7.60$\pm$0.08  &    2	 &   7.60$\pm$0.04  & 0.00   \\
Si\,{\sc i}  & 7.28$\pm$0.09 & 0.13$\pm$0.14&  0.13$\pm$0.19 & 11  &  7.50$\pm$0.07  &    12 &   7.51$\pm$0.03  & 0.01  \\
Ca\,{\sc i}  & 6.09$\pm$0.10 & 0.10$\pm$0.15&  0.10$\pm$0.19 & 18  &  6.34$\pm$0.08  &    18 &   6.34$\pm$0.04  & 0.00   \\
Sc\,{\sc i}  & 2.78$\pm$0.00 & 0.01$\pm$0.08& -0.33$\pm$0.14 & 1   &  3.12$\pm$0.00  &    1	 &   3.15$\pm$0.04  & 0.03  \\
Sc\,{\sc ii} & 3.05$\pm$0.06 & 0.17$\pm$0.13&  0.18$\pm$0.20 & 7   &  3.23$\pm$0.08  &    7	 &   3.15$\pm$0.04  & 0.08   \\
Ti\,{\sc i}  & 4.82$\pm$0.12 & 0.21$\pm$0.17&  0.21$\pm$0.21 & 39  &  4.96$\pm$0.09  &    43 &   4.95$\pm$0.05  & 0.01   \\
Ti\,{\sc ii} & 4.86$\pm$0.13 & 0.22$\pm$0.17&  0.18$\pm$0.19 & 12  &  4.99$\pm$0.08  &    12 &   4.95$\pm$0.05  & 0.04   \\
V\,{\sc i}   & 3.81$\pm$0.07 & 0.17$\pm$0.12&  0.05$\pm$0.15 & 4   &  3.99$\pm$0.05  &    5	 &   3.93$\pm$0.08  & 0.06   \\
Cr\,{\sc i}  & 5.30$\pm$0.11 &-0.06$\pm$0.15& -0.01$\pm$0.18 & 19  &  5.71$\pm$0.07  &    19 &   5.64$\pm$0.04  & 0.07   \\
Cr\,{\sc ii} & 5.32$\pm$0.16 & 0.03$\pm$0.23& -0.01$\pm$0.20 & 3   &  5.64$\pm$0.14  &    3	 &   5.64$\pm$0.04  & 0.00   \\
Mn\,{\sc i}  & 5.06$\pm$0.15 &-0.21$\pm$0.21& -0.23$\pm$0.23 & 12  &  5.62$\pm$0.13  &    13 &   5.43$\pm$0.05  & 0.19   \\
Fe\,{\sc i}  & 7.17$\pm$0.12 &-0.02$\pm$0.17& -0.01$\pm$0.20 & 128 &  7.54$\pm$0.09  &    132&   7.50$\pm$0.04  & 0.04   \\
Fe\,{\sc ii} & 7.16$\pm$0.07 & 0.00$\pm$0.11&  0.00$\pm$0.19 & 15  &  7.51$\pm$0.04  &    17 &   7.50$\pm$0.04  & 0.01   \\
Co\,{\sc i}  & 4.62$\pm$0.15 & 0.00$\pm$0.23&  0.03$\pm$0.26 & 6   &  4.97$\pm$0.15  &    7	 &   4.99$\pm$0.07  & 0.02  \\
Ni\,{\sc i}  & 5.91$\pm$0.12 &-0.02$\pm$0.17& -0.03$\pm$0.21 & 50  &  6.28$\pm$0.09  &    54 &   6.22$\pm$0.04  & 0.06   \\
Zn\,{\sc i}  & 4.58$\pm$0.01 & 0.25$\pm$0.09&  0.26$\pm$0.17 & 2   &  4.68$\pm$0.03  &    2	 &   4.56$\pm$0.05  & 0.12   \\
Sr\,{\sc i}  & 2.28$\pm$0.00 &-0.28$\pm$0.08& -0.16$\pm$0.14 & 1   &  2.91$\pm$0.00  &    1	 &   2.87$\pm$0.07  & 0.04   \\
Y\,{\sc ii}  & 1.92$\pm$0.32 &-0.02$\pm$0.33& -0.15$\pm$0.22 & 2   &  2.29$\pm$0.05  &    2	 &   2.21$\pm$0.05  & 0.08   \\
Zr\,{\sc ii} & 2.45$\pm$0.00 & 0.12$\pm$0.08&  0.27$\pm$0.14 & 1   &  2.68$\pm$0.00  &    1	 &   2.58$\pm$0.04  & 0.10   \\
Ba\,{\sc ii} & 1.88$\pm$0.09 &-0.01$\pm$0.13&  0.06$\pm$0.20 & 4   &  2.24$\pm$0.06  &    4	 &   2.18$\pm$0.09  & 0.06   \\
Ce\,{\sc ii} & 1.40$\pm$0.09 & 0.11$\pm$0.12&  0.25$\pm$0.21 & 2   &  1.64$\pm$0.02  &    2	 &   1.58$\pm$0.04  & 0.06   \\
Nd\,{\sc ii} & 1.08$\pm$0.00 & 0.01$\pm$0.09&  0.19$\pm$0.14 & 1   &  1.42$\pm$0.05  &    3	 &   1.42$\pm$0.04  & 0.00   \\
Sm\,{\sc ii} &  --	         & --   	    &         --     &  -- &  0.96$\pm$0.00  &    1	 &   0.96$\pm$0.04  & 0.00   \\
\hline
\end{tabular}
\small
\\
(1) The abundances are obtained using the {\sc ELODIE} spectrum. (2) The abundances are obtained using the {\sc PolarBASE} spectrum. (*) $\Delta {\rm log} \epsilon_{\rm \odot}{\rm (X) = log} \epsilon_{\rm \odot} {\rm (X)}_{\rm This\,study} - {\rm log} \epsilon_{\rm \odot} {\rm (X)}_{\rm Asplund}$. ($\dagger$) The Solar abundances calculated in this study.
\end{table*}

Further verification of the $\log gf$ values was performed by comparing the $\log gf$ values used in this study with those in the {\it Gaia}-ESO line list v.6 provided by the GES collaboration \citep{heiter2021}. It should be noted that the $gf$-values for the chosen lines of Fe\,{\sc i} and Fe\,{\sc ii} in this study were taken from the compilation of \citet {fuhr2006}. The GES line list contains a list of recommended lines and atomic data (hyperfine structure-has-corrected $gf$-values) for the analysis of FGK stars. It is worth noting that several lines in the spectra of FGK stars have not yet been identified \citep{heiter2015}.  

The GES line list (v.6) contains a total of 141\,233 lines in the wavelength range 4200 – 10\,000 \AA. This number is reduced to 84\,735 lines in the wavelength range 4200 - 6800 \AA. For species that match those in the line list (24 in total) provided in this study in the same wavelength range, the GES contains 50\,326 lines. This wavelength range is the same as the wavelength range for the {\sc ELODIE} spectrum used for diagnostics. Comparing the GES line list with the new line list provided useful information. We used a total of 363 lines from the line list to match the GES lines. Out of 363 lines, 11 lines were found to be out of range ($<$ 4200 \AA) in the GES line list and the two were missing (Y\,{\sc ii} lines at 4883.69 \AA\, and 5087.43 \AA). The search resulted in a one-to-one match of 350 lines in terms of wavelength, element, $\log gf$ and low LEP. The difference between the $\log gf$ values\footnote{$\log gf_{\rm This\, study}$ - $\log gf_{\rm GES}$} for the 128 common Fe\,{\sc i} lines in the GES line list was -0.03$\pm$0.06 dex. For the 16 Fe\,{\sc ii} lines, the difference in $\log gf$ values was -0.002$\pm$0.079 dex. For, Na\,{\sc i}, Si\,{\sc i}, Ca\,{\sc i}, Sc\,{\sc ii}, Ti\,{\sc i}, Ti\,{\sc ii}, V\,{\sc i}, Cr\,{\sc i}, Ni\,{\sc i}, Zn\,{\sc i}, Sr\,{\sc i}, Zr\,{\sc ii}, and Sm\,{\sc ii}, the difference in the $\log gf$ values was less than 0.05 dex. For 54 common Ni\,{\sc i} lines, the difference was found to be -0.045$\pm$0.114 dex. A formal least square solution gives a gradient of 0.99 and a zero point difference of -0.02 dex over 350 common lines. However, there were large discrepancies in the $\log gf$ values for transitions for some elements common to both line lists. Details about this discrepancy are included in the following section.

\subsection{Model Parameters and Abundances}

We used neutral and ionised Fe lines to determine model atmospheric parameters such as effective temperature, surface gravity, microturbulence, and metallicity (Table \ref{tab:model_param}). First, the effective temperature was calculated by requiring that the resulting abundance be independent of the lower LEP.

\begin{figure*}
\centering
\includegraphics[width=\columnwidth]{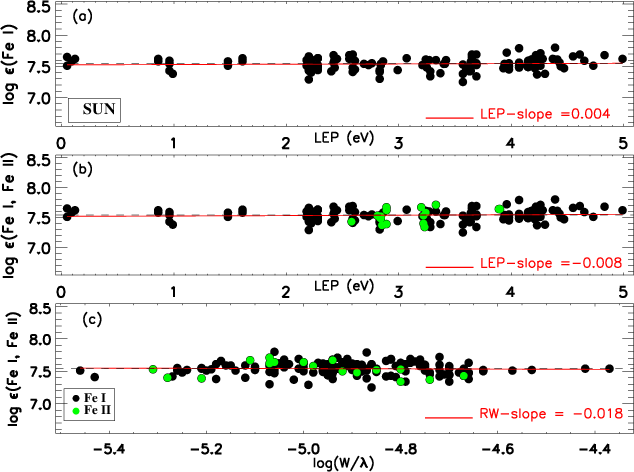}
\includegraphics[width=\columnwidth]{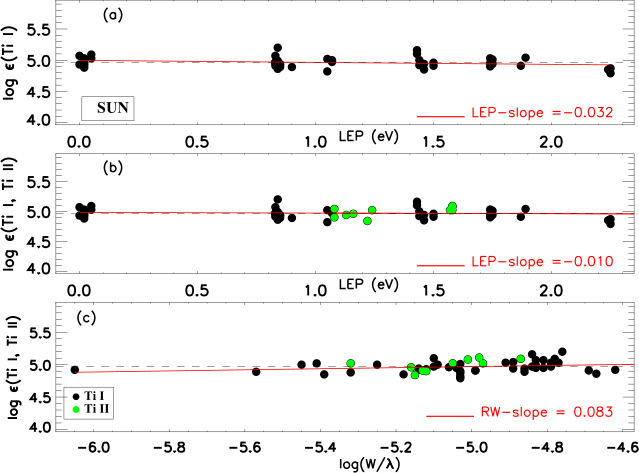}
\caption{An example for the determination of the atmospheric parameters $T_{\rm eff}$ and $\xi$ using abundance ($\log\epsilon$) as a function of both lower LEP (panels a and b for Fe and Ti, respectively) and reduced EW (REW; $\log$ (EW/$\lambda$), panels c for Fe and Ti, respectively) for the Sun. In all panels, the solid red line represents the least-squares fit to the data.}
\label{fig:sun_model_param}
\end{figure*}

\begin{figure*}
\centering
\includegraphics[width=\columnwidth]{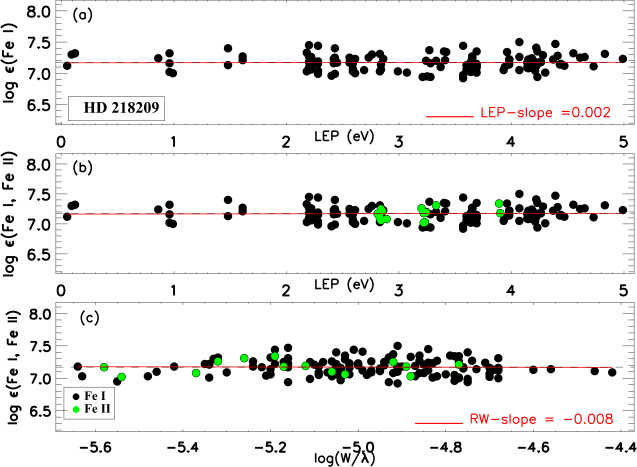}
\includegraphics[width=\columnwidth]{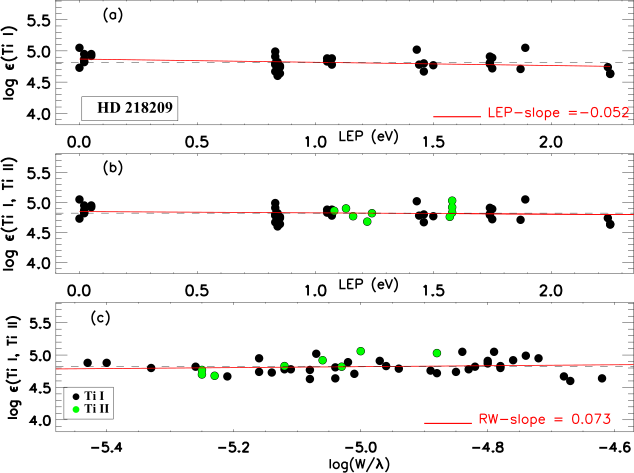}
\caption{An example for the determination of the atmospheric parameters $T_{\rm eff}$ and $\xi$ using abundance ($\log\epsilon$) as a function of both lower LEP (panels a and b for Fe and Ti, respectively) and reduced EW (REW; $\log$ (EW/$\lambda$), panels c for Fe and Ti, respectively) for HD\,218209 and the {\sc ELODIE} spectrum. In all panels, the solid red line represents the least-squares fit to the data.}
\label{fig:hd_model_param}
\end{figure*}

\begin{table}
\setlength{\tabcolsep}{4pt}
\renewcommand{\arraystretch}{1.5}
\caption{Model atmosphere parameters for HD\,218209 and the Sun.}
\label{tab:model_param}
\centering
\begin{tabular}{lcccc}
\hline
Star	&	$T_{\rm eff}$	&	$\log g$ 	&	[Fe/H] &  $\xi$ \\
    & (K) & (cgs) & (dex) & (km s$^{\rm -1}$) \\
    \hline
HD\,218209$^{\rm \dagger}$ & 5650$_{-170}^{+165}$ & 4.55$_{-0.21}^{+0.18}$ & -0.35	$_{-0.07}^{+0.07}$ & 0.60 $_{-0.50}^{+0.50}$ \\
HD\,218209$^{\rm *}$ & 	5630$_{-160}^{+170}$ & 4.43$_{-0.26}^{+0.22}$ & -0.31	$_{-0.13}^{+0.13}$  & 0.35 $_{-0.50}^{+0.50}$       \\
Sun        & 5790$_{-125}^{+125}$ &	4.40$_{-0.23}^{+0.16}$	&  0.00	$_{-0.04}^{+0.04}$ & 0.66 $_{-0.50}^{+0.50}$ \\
\hline
\end{tabular}
\begin{list}{}{}
\item ($\dagger$) The ELODIE spectrum was used.
\item (*) The PolarBase spectrum was used.
\end{list}
\end{table}
 If all lines have the same LEP and a similar wavelength, the microturbulence ($\xi$) is determined by requiring that the calculated abundance be independent of the reduced equivalent width (EW). The precision in the determination of the microturbulent velocity is $\pm$0.5 km s$^{-1}$. We calculated surface gravity ($\log g$) by requiring ionisation equilibrium, which means that the Fe\,{\sc i} and Fe\,{\sc ii} lines yield the same iron abundance (Figures \ref{fig:sun_model_param} and \ref{fig:hd_model_param}). Due to the interdependence of model parameters, an iterative procedure is required. Between each of the above steps, minor adjustments are made to the model parameters. We also confirmed that there is no substantial trend in iron abundances (see Figures \ref{fig:sun_model_param} and \ref{fig:hd_model_param} for the Sun and HD\,218209, respectively). 

The model parameters obtained for the Sun as a result of the Solar analysis were determined as $T_{\rm eff}= 5790$ K, $\log g = 4.4$ cgs, [Fe/H]=0 dex and $\xi= 0.66$ km s$^{\rm -1}$. These values are similar to those recommended by \citet{heiter2015} as ($T_{\rm eff}$, $\log g$) = (5771, 4.438). Element abundances for the Sun were calculated with these model parameters. For HD\,218209, the model parameters were obtained from {\sc ELODIE} and {\sc PolarBASE} spectrum of the star. 

During the calculation of the model atmosphere, the effect of convection on the reported abundances is also taken into consideration. In one-dimensional (1D) atmosphere modeling, it is currently common practice to assume a universal value of 1.5 for the mixing length parameter, $\alpha$. However, in order to test the impact of convection on final abundances in this study, the ATLAS models were produced using two different methods for calculating $\alpha$. Two well known references in the literature, \citet{Ludwig1999} and \citet{Magic2015}, for the calculation of $\alpha$, yields two different values (1.59 and 1.98) even when the same set of model parameters is employed. The logarithmic abundance of ionized iron showed a 0.01 dex increase for the model atmosphere computed with the $\alpha$ parameter set to 1.59 and the model parameters from the {\sc ELODIE} spectrum. For $\alpha=$1.98, it is +0.02 dex. With regards to the FGK dwarf field stars, it is expected that the influence of the mixing length parameter, for instance, on the determination of metallicity would be minimal, amounting to less than 0.02 dex over eight dwarf stars \citep{Song2020}. Thus, our results for HD\,218209 confirm those of \citet{Song2020}. Regarding the model parameters of HD\,218209 from the {\sc PolarBASE} spectrum, the logarithmic abundance of ionized iron remained the same for the model atmosphere computed with $\alpha=$1.59. The difference in logarithmic abundance is 0.02 dex for $\alpha=$1.98.

The resulting stellar model parameters for HD\,218209 along with our determination of model parameters of the Sun are listed in Table \ref{tab:model_param}.  Table \ref{tab:hd218209_model_param_lit} lists the model parameters reported for the star in the literature. \citet{Rebolo1988}, \citet{Abia1988}, and \citet{Kim2016} show the largest variations compared to our $\log g$. The abundances obtained for the Solar photosphere as a result of the Solar analysis are given in Table \ref{tab:solar_abund}. The Solar abundances from \citet{asplund2009} are also included for comparison. In Table \ref{tab:solar_abund}, we provide a summary of element abundances based on LTE-based model parameters. $\log \epsilon$ is the logarithm of the abundances. The errors reported in $\log \epsilon$ abundances are the 1$\sigma$ line-to-line scatter in abundances. [X/H] is the logarithmic abundance ratio of hydrogen to the corresponding Solar value, and [X/Fe] is the logarithmic abundance considering the Fe\,{\sc i} abundance. The error in [X/Fe] is the square root of the sum of the quadratures of the errors in [X/H] and [Fe/H]. Table \ref{tab:solar_abund} also presents the abundances obtained using the PolarBase spectrum of the star as [X/Fe] ratio.

\begin{table}
\setlength{\tabcolsep}{2pt}
\renewcommand{\arraystretch}{1.2}
\caption{Atmospheric parameters of HD\,218209 from this study.}
\label{tab:hd218209_model_param_lit}
\centering
\begin{tabular}{lccl}
\hline
$T_{\rm eff}$	&	$\log g$ 	&	[Fe/H] & ~~~~~~~~~~~~~~~~~Notes\\
(K)    & (cgs)    & (dex)  & \\
\hline
5650 & 4.55 & -0.37  & This study ({\sc ELODIE})\\
5630 & 4.43 & -0.32  & This study ({\sc PolarBASE})\\
5506 & 4.38 & -0.46  & \citet{takeda2023}\\
5636 & 4.56 & -0.47  & \citet{Rice2020}\\
5518 & 4.39 & -0.49  & \citet{Aguilera-Gómez2018} \\
5623 & 4.46 & -0.40  & \citet{Luck2017}\\
5555 & 4.26 & -0.49  & \citet{Boeche2016} \\
5565 & 4.29 & -0.49  & \citet{Boeche2016}\\
5669 & 4.24 & -0.39  & \citet{Kim2016} \\
5575 & 4.11 & -0.46  & \citet{Kim2016} \\
5607 & 4.07 & -0.58  & \citet{Kim2016} \\  
5536 & 4.37 & -0.46  & \citet{DaSilva2015}  \\
5705 & 4.50 & -0.43  & \citet{Mishenina2015} \\
5705 & 4.50 & -0.43  & \citet{Mishenina2013} \\
5592 & 4.25 & -0.51  & \citet{Lee2011} \\
5592 & 4.33 & -0.61  & \citet{Lee2011} \\
5592 & 4.31 & -0.65  & \citet{Lee2011} \\
5705 & 4.50 & -0.43  & \citet{Mishenina2011} \\
5539 & 4.37 & -0.50  & \citet{DaSilva2011} \\
5600 & --   & --     & \citet{Casagrande2011} \\
5473 & 4.57 & -0.64  & \citet{Sozzetti2009} \\
5506 & 4.38 & -0.46  & \citet{Takeda2007} \\
5693 & --   & --     & \citet{Masana2006}\\
5585 & 4.60 & -0.46  & \citet{Valenti2005} \\
5684 & --   & --     & \citet{Kovtyukh2004} \\
5705 & 4.50 & -0.43  & \citet{Mishenina2004} \\
5665 & 4.40 & -0.60  & \citet{Gehren2004} \\
5478 & 4.00 & -0.60  & \citet{Rebolo1988} \\
5478 & 4.00 & -0.60  & \citet{Abia1988}  \\
\hline
\end{tabular}
\end{table}

\section{Summary and Conclusion}

We present a line list that can be useful for abundance analysis of G-type stars over 4080 -- 6780 \AA\, wavelength region (i.e. for spectra from the {\sc ELODIE}/{\sc SOPHIE} library). The line list is expected to be useful for the surveys/libraries with overlapping spectral regions (e.g. {\sc ELODIE} library, UVES-580 setting of {\it Gaia}-ESO, and, especially, for the analysis of F- and G-type stars in general. We identified 363 lines of 24 species that have accurate $gf$-values and are free of blends in the spectra of the Sun and a Solar analogue star, HD\,218209. It should be noted that the GES line list does not contain any atomic transitions below 4200 \AA\, and unlike the GES line list, the line list created in this study contains additional four Fe\,{\sc i}, one Fe\,{\sc ii}, 1 Sc\,{\sc i}, 2 Ti\,{\sc i} and 1 Co\,{\sc i} line below this limit. To assess the uncertainty in $\log gf$-values and to minimise systematic errors, we calculated Solar abundances using stellar lines and the high-resolution KPNO spectrum of the Sun. For the creation of the line list in this study, high resolution and high signal to noise {\sc ELODIE} spectrum of HD\,218209 was used. During the comparison with the GES line list, discrepancies in the $\log gf$ values of some ions were detected. For example, the average difference in $\log gf$ values for 5 V\,{\sc i} lines was 0.325 +- 0.328. Similarly, for 11 Mn\,{\sc i} lines, the difference was 0.288 +- 0.283 dex. The largest difference detected was 0.585$\pm$0.161 dex over 6 Co\,{\sc i} lines. For the other elements reported in the new line list presented in this study, the $\log gf$ difference was less than 0.1 dex.

To verify the accuracy of the model parameters obtained for HD\,218209 from the {\sc ELODIE} spectrum, we analyzed the {\sc PolarBASE} spectrum of the same star. The model parameters from both spectra matched quite well, taking into account the error margins. The element abundances obtained from both spectra also matched quite well, except for Sc{\sc i} which was represented by a single line at 4023.69 \AA\ in both spectra. The logarithmic abundance from this line in the {\sc ELODIE} showed a 0.3 dex increase. Additionally, three other species (Sr\, {\sc i}, Zr\, {\sc ii}, and Ce\, {\sc ii}), with the latter two species presented by two transitions in the spectra, had discrepancies in their logarithmic abundances at around 0.2 dex level. Although the {\sc PolarBASE} spectrum had higher resolution than the {\sc ELODIE}, several lines in the {\sc PolarBASE} did not have clear line profiles. Unfortunately, there was only one spectrum available in the {\sc PolarBASE} archive, so a more thorough analysis is recommended to check the abundances of the star from the {\sc PolarBASE} spectrum.

The model parameters obtained for HD\,218209 clearly indicate a Solar analogue character for the star. Although no definite classification scheme exists, Solar analogue stars are defined as the dwarf stars of early G-type which have properties analogous to the Sun (e.g. differences in $T_{\rm eff}$ and $\log g$ are $\pm \leq$ 100–200 K and $\pm \leq$ 0.1–0.2 dex). Indeed, seeing the star listed in a recent study by \citet{takeda2023} as a Solar analogue was not surprising for us. The model parameters ($T_{\rm eff}=5506$ K, $\log g=4.38$ cgs, [Fe/H]=-0.46 dex, $\xi=$ 0.74 km s$^{\rm -1}$) reported by \citet{takeda2023} are in good agreement with those obtained in this study. However, we report up-to-date abundances for the star using 24 species identified in both {\sc ELODIE} and {\sc PolarBase} spectra of the star.

HD\,218209 is listed in {\it Gaia} DR2  \citep{gaia2018} with {\it Gaia} DR2 2213415145505277824 designation and a $T_{\rm eff}$ of 5648 K. The surface gravity and metallicity of the star were not reported. HD\,218209 is also listed in the {\it Gaia} DR3 \citep{gaia2023}. Our spectroscopic measurements of the temperature and surface gravity of the star align very well with the values reported by the {\it Gaia} consortium, which are $T_{\rm eff}=5528$ K and $\log g = 4.40$ cgs. However, there seems to be an error in the reported metallicity of the star by {\it Gaia}, which states [Fe/H] = -0.79 dex. This discrepancy is likely due to the incorrect selection of the template spectrum for the star in the {\it Gaia} DR3, where the metallicity of the template star is reported as -0.50 dex.

In the {\it Gaia} DR3 \citep{gaia2023}, an important addition is a collection of 220 million low-resolution BP/RP spectra. The GSP-Phot catalogue provides consistent estimations of stellar parameters for 471 million sources with $G<19$ mag, derived from the BP/RP spectra, parallax, and integrated photometry. It assumes that each source is a single star and that any intrinsic time variability is lost when using combined BP/RP spectra. However, it is important to note that the GSP-Phot results from the BP/RP spectra in the {\it Gaia} DR3 may still be affected by systematic effects, particularly in the [M/H] estimates with large systematic errors. Consequently, it is advised not to rely on these estimates as they can only provide qualitative information at best \citep{Beck2023}.

In a study of stellar and sub-stellar companions of nearby stars from the {\it Gaia} DR2 \citep{gaia2018}, \citet{Kervella2019} examined the status of stars exhibiting anomalies in their proper motions as binary stars. HD \,218209 as a high proper-motion star happens to be one of the common stars. The study found no evidence of a companion around HD\,218209, even at distances ranging from 1 to 50 astronomical units (au) and down to planetary masses. Therefore, the authors concluded that HD\,218209 is a single star (P. Kervella, private communication, 2023). 

\citet{soubiran2005}, in their exploration of abundance discrepancies between the thin disc and thick disc of the Galaxy, identified HD\,218209 out of the 44 stars as being part of the Hercules stream with a likelihood of 0.76. The Hercules stream shares chemical characteristics with the thin disc, which bolsters the dynamic hypothesis (i.e. the influence of the central bar of the Galaxy impacting  stars) of its origin. However, kinematic and orbital dynamics analysis for the star is the subject of another study.

\section*{Acknowledgements}
This study has partly been supported by the Scientific and Technological Research Council (TÜBİTAK) MFAG-121F265. We thank Ferhat Güney for his help with the preparation of Figure 2. We also thank Gizay YOLALAN and Sena A. \c{S}ENT\"{Ü}RK for helpful discussion.  

\bibliographystyle{mnras}
\bibliography{references}

\begin{thebibliography}{}
\makeatletter
\relax
\def\mn@urlcharsother{\let\do\@makeother \do\$\do\&\do\#\do\^\do\_\do\%\do\~}
\def\mn@doi{\begingroup\mn@urlcharsother \@ifnextchar [ {\mn@doi@}
  {\mn@doi@[]}}
\def\mn@doi@[#1]#2{\def\@tempa{#1}\ifx\@tempa\@empty \href
  {http://dx.doi.org/#2} {doi:#2}\else \href {http://dx.doi.org/#2} {#1}\fi
  \endgroup}
\def\mn@eprint#1#2{\mn@eprint@#1:#2::\@nil}
\def\mn@eprint@arXiv#1{\href {http://arxiv.org/abs/#1} {{\tt arXiv:#1}}}
\def\mn@eprint@dblp#1{\href {http://dblp.uni-trier.de/rec/bibtex/#1.xml}
  {dblp:#1}}
\def\mn@eprint@#1:#2:#3:#4\@nil{\def\@tempa {#1}\def\@tempb {#2}\def\@tempc
  {#3}\ifx \@tempc \@empty \let \@tempc \@tempb \let \@tempb \@tempa \fi \ifx
  \@tempb \@empty \def\@tempb {arXiv}\fi \@ifundefined
  {mn@eprint@\@tempb}{\@tempb:\@tempc}{\expandafter \expandafter \csname
  mn@eprint@\@tempb\endcsname \expandafter{\@tempc}}}

\bibitem[\protect\citeauthoryear{{Abia}, {Rebolo}, {Beckman}  \&
  {Crivellari}}{{Abia} et~al.}{1988}]{Abia1988}
{Abia} C.,  {Rebolo} R.,  {Beckman} J.~E.,   {Crivellari} L.,  1988, \aap,
  \href {https://ui.adsabs.harvard.edu/abs/1988A&A...206..100A} {206, 100}

\bibitem[\protect\citeauthoryear{{Aguilera-G{\'o}mez}, {Ram{\'\i}rez}  \&
  {Chanam{\'e}}}{{Aguilera-G{\'o}mez} et~al.}{2018}]{Aguilera-Gómez2018}
{Aguilera-G{\'o}mez} C.,  {Ram{\'\i}rez} I.,   {Chanam{\'e}} J.,  2018, \mn@doi
  [\aap] {10.1051/0004-6361/201732209}, \href
  {https://ui.adsabs.harvard.edu/abs/2018A&A...614A..55A} {614, A55}

\bibitem[\protect\citeauthoryear{{Allende Prieto} et~al.,}{{Allende Prieto}
  et~al.}{2008}]{allende2008}
{Allende Prieto} C.,  et~al., 2008, \mn@doi [Astronomische Nachrichten]
  {10.1002/asna.200811080}, \href
  {https://ui.adsabs.harvard.edu/abs/2008AN....329.1018A} {329, 1018}

\bibitem[\protect\citeauthoryear{{Asplund}, {Grevesse}, {Sauval}  \&
  {Scott}}{{Asplund} et~al.}{2009}]{asplund2009}
{Asplund} M.,  {Grevesse} N.,  {Sauval} A.~J.,   {Scott} P.,  2009, \mn@doi
  [\araa] {10.1146/annurev.astro.46.060407.145222}, \href
  {https://ui.adsabs.harvard.edu/abs/2009ARA&A..47..481A} {47, 481}

\bibitem[\protect\citeauthoryear{{Bard} \& {Kock}}{{Bard} \&
  {Kock}}{1994}]{Bard1994}
{Bard} A.,  {Kock} M.,  1994, \aap, \href
  {https://ui.adsabs.harvard.edu/abs/1994A&A...282.1014B} {282, 1014}

\bibitem[\protect\citeauthoryear{{Beck} et~al.,}{{Beck}
  et~al.}{2023}]{Beck2023}
{Beck} P.~G.,  et~al., 2023, \mn@doi [arXiv e-prints]
  {10.48550/arXiv.2307.10812}, \href
  {https://ui.adsabs.harvard.edu/abs/2023arXiv230710812B} {p. arXiv:2307.10812}

\bibitem[\protect\citeauthoryear{{Biemont} \& {Godefroid}}{{Biemont} \&
  {Godefroid}}{1980}]{Biemont1980}
{Biemont} E.,  {Godefroid} M.,  1980, \aap, \href
  {https://ui.adsabs.harvard.edu/abs/1980A&A....84..361B} {84, 361}

\bibitem[\protect\citeauthoryear{{Biemont}, {Grevesse}, {Hannaford}  \&
  {Lowe}}{{Biemont} et~al.}{1981}]{Biemont1981}
{Biemont} E.,  {Grevesse} N.,  {Hannaford} P.,   {Lowe} R.~M.,  1981, \mn@doi
  [\apj] {10.1086/159213}, \href
  {https://ui.adsabs.harvard.edu/abs/1981ApJ...248..867B} {248, 867}

\bibitem[\protect\citeauthoryear{{Boeche} \& {Grebel}}{{Boeche} \&
  {Grebel}}{2016}]{Boeche2016}
{Boeche} C.,  {Grebel} E.~K.,  2016, \mn@doi [\aap]
  {10.1051/0004-6361/201526758}, \href
  {https://ui.adsabs.harvard.edu/abs/2016A&A...587A...2B} {587, A2}

\bibitem[\protect\citeauthoryear{{Casagrande}, {Sch{\"o}nrich}, {Asplund},
  {Cassisi}, {Ram{\'\i}rez}, {Mel{\'e}ndez}, {Bensby}  \&
  {Feltzing}}{{Casagrande} et~al.}{2011}]{Casagrande2011}
{Casagrande} L.,  {Sch{\"o}nrich} R.,  {Asplund} M.,  {Cassisi} S.,
  {Ram{\'\i}rez} I.,  {Mel{\'e}ndez} J.,  {Bensby} T.,   {Feltzing} S.,  2011,
  \mn@doi [\aap] {10.1051/0004-6361/201016276}, \href
  {https://ui.adsabs.harvard.edu/abs/2011A&A...530A.138C} {530, A138}

\bibitem[\protect\citeauthoryear{{Castelli} \& {Kurucz}}{{Castelli} \&
  {Kurucz}}{2003}]{Castelli2003}
{Castelli} F.,  {Kurucz} R.~L.,  2003, in {Piskunov} N.,  {Weiss} W.~W.,
  {Gray} D.~F.,  eds,  Published on behalf of the IAU by the Astronomical
  Society of the Pacific Vol. 210, Modelling of Stellar Atmospheres. p.~A20
  (\mn@eprint {arXiv} {astro-ph/0405087}),
  \mn@doi{10.48550/arXiv.astro-ph/0405087}

\bibitem[\protect\citeauthoryear{{Da Silva}, {Milone}  \& {Reddy}}{{Da Silva}
  et~al.}{2011}]{DaSilva2011}
{Da Silva} R.,  {Milone} A.~C.,   {Reddy} B.~E.,  2011, \mn@doi [\aap]
  {10.1051/0004-6361/201015907}, \href
  {https://ui.adsabs.harvard.edu/abs/2011A&A...526A..71D} {526, A71}

\bibitem[\protect\citeauthoryear{{Da Silva}, {Milone}  \& {Rocha-Pinto}}{{Da
  Silva} et~al.}{2015}]{DaSilva2015}
{Da Silva} R.,  {Milone} A. d.~C.,   {Rocha-Pinto} H.~J.,  2015, \mn@doi [\aap]
  {10.1051/0004-6361/201525770}, \href
  {https://ui.adsabs.harvard.edu/abs/2015A&A...580A..24D} {580, A24}

\bibitem[\protect\citeauthoryear{{De Silva} et~al.,}{{De Silva}
  et~al.}{2015}]{silva2015}
{De Silva} G.~M.,  et~al., 2015, \mn@doi [\mnras] {10.1093/mnras/stv327}, \href
  {https://ui.adsabs.harvard.edu/abs/2015MNRAS.449.2604D} {449, 2604}

\bibitem[\protect\citeauthoryear{{Den Hartog}, {Lawler}, {Sneden}  \&
  {Cowan}}{{Den Hartog} et~al.}{2003}]{DenHartog2003}
{Den Hartog} E.~A.,  {Lawler} J.~E.,  {Sneden} C.,   {Cowan} J.~J.,  2003,
  \mn@doi [\apjs] {10.1086/376940}, \href
  {https://ui.adsabs.harvard.edu/abs/2003ApJS..148..543D} {148, 543}

\bibitem[\protect\citeauthoryear{{Den Hartog}, {Lawler}, {Sobeck}, {Sneden}  \&
  {Cowan}}{{Den Hartog} et~al.}{2011}]{DenHartog2011}
{Den Hartog} E.~A.,  {Lawler} J.~E.,  {Sobeck} J.~S.,  {Sneden} C.,   {Cowan}
  J.~J.,  2011, \mn@doi [\apjs] {10.1088/0067-0049/194/2/35}, \href
  {https://ui.adsabs.harvard.edu/abs/2011ApJS..194...35D} {194, 35}

\bibitem[\protect\citeauthoryear{{Den Hartog}, {Lawler}, {Sneden}, {Cowan},
  {Roederer}  \& {Sobeck}}{{Den Hartog} et~al.}{2021}]{DenHartog2021}
{Den Hartog} E.~A.,  {Lawler} J.~E.,  {Sneden} C.,  {Cowan} J.~J.,  {Roederer}
  I.~U.,   {Sobeck} J.,  2021, \mn@doi [\apjs] {10.3847/1538-4365/ac04b1},
  \href {https://ui.adsabs.harvard.edu/abs/2021ApJS..255...27D} {255, 27}

\bibitem[\protect\citeauthoryear{{Fuhr} \& {Wiese}}{{Fuhr} \&
  {Wiese}}{2006}]{fuhr2006}
{Fuhr} J.~R.,  {Wiese} W.~L.,  2006, \mn@doi [Journal of Physical and Chemical
  Reference Data] {10.1063/1.2218876}, \href
  {https://ui.adsabs.harvard.edu/abs/2006JPCRD..35.1669F} {35, 1669}

\bibitem[\protect\citeauthoryear{{Gaia Collaboration} et~al.,}{{Gaia
  Collaboration} et~al.}{2018}]{gaia2018}
{Gaia Collaboration} et~al., 2018, \mn@doi [\aap]
  {10.1051/0004-6361/201832865}, \href
  {https://ui.adsabs.harvard.edu/abs/2018A&A...616A..11G} {616, A11}

\bibitem[\protect\citeauthoryear{{Gaia Collaboration} et~al.,}{{Gaia
  Collaboration} et~al.}{2023}]{gaia2023}
{Gaia Collaboration} et~al., 2023, \mn@doi [\aap]
  {10.1051/0004-6361/202243940}, \href
  {https://ui.adsabs.harvard.edu/abs/2023A&A...674A...1G} {674, A1}

\bibitem[\protect\citeauthoryear{{Gehren}, {Liang}, {Shi}, {Zhang}  \&
  {Zhao}}{{Gehren} et~al.}{2004}]{Gehren2004}
{Gehren} T.,  {Liang} Y.~C.,  {Shi} J.~R.,  {Zhang} H.~W.,   {Zhao} G.,  2004,
  \mn@doi [\aap] {10.1051/0004-6361:20031582}, \href
  {https://ui.adsabs.harvard.edu/abs/2004A&A...413.1045G} {413, 1045}

\bibitem[\protect\citeauthoryear{{Gilmore} et~al.,}{{Gilmore}
  et~al.}{2012}]{gilmore2012}
{Gilmore} G.,  et~al., 2012, The Messenger, \href
  {https://ui.adsabs.harvard.edu/abs/2012Msngr.147...25G} {147, 25}

\bibitem[\protect\citeauthoryear{{Hannaford}, {Lowe}, {Grevesse}, {Biemont}  \&
  {Whaling}}{{Hannaford} et~al.}{1982}]{Hannaford1982}
{Hannaford} P.,  {Lowe} R.~M.,  {Grevesse} N.,  {Biemont} E.,   {Whaling} W.,
  1982, \mn@doi [\apj] {10.1086/160384}, \href
  {https://ui.adsabs.harvard.edu/abs/1982ApJ...261..736H} {261, 736}

\bibitem[\protect\citeauthoryear{{Hansen}, {Bergemann}, {Cescutti},
  {Fran{\c{c}}ois}, {Arcones}, {Karakas}, {Lind}  \& {Chiappini}}{{Hansen}
  et~al.}{2013}]{Hansen2013}
{Hansen} C.~J.,  {Bergemann} M.,  {Cescutti} G.,  {Fran{\c{c}}ois} P.,
  {Arcones} A.,  {Karakas} A.~I.,  {Lind} K.,   {Chiappini} C.,  2013, \mn@doi
  [\aap] {10.1051/0004-6361/201220584}, \href
  {https://ui.adsabs.harvard.edu/abs/2013A&A...551A..57H} {551, A57}

\bibitem[\protect\citeauthoryear{{Heijmans} et~al.,}{{Heijmans}
  et~al.}{2012}]{heijmans2012}
{Heijmans} J.,  et~al., 2012, in {McLean} I.~S.,  {Ramsay} S.~K.,   {Takami}
  H.,  eds,  Society of Photo-Optical Instrumentation Engineers (SPIE)
  Conference Series Vol. 8446, Ground-based and Airborne Instrumentation for
  Astronomy IV. p. 84460W, \mn@doi{10.1117/12.925806}

\bibitem[\protect\citeauthoryear{{Heiter}, {Jofr{\'e}}, {Gustafsson}, {Korn},
  {Soubiran}  \& {Th{\'e}venin}}{{Heiter} et~al.}{2015}]{heiter2015}
{Heiter} U.,  {Jofr{\'e}} P.,  {Gustafsson} B.,  {Korn} A.~J.,  {Soubiran} C.,
   {Th{\'e}venin} F.,  2015, \mn@doi [\aap] {10.1051/0004-6361/201526319},
  \href {https://ui.adsabs.harvard.edu/abs/2015A&A...582A..49H} {582, A49}

\bibitem[\protect\citeauthoryear{{Heiter} et~al.,}{{Heiter}
  et~al.}{2021}]{heiter2021}
{Heiter} U.,  et~al., 2021, \mn@doi [\aap] {10.1051/0004-6361/201936291}, \href
  {https://ui.adsabs.harvard.edu/abs/2021A&A...645A.106H} {645, A106}

\bibitem[\protect\citeauthoryear{{Kervella}, {Arenou}, {Mignard}  \&
  {Th{\'e}venin}}{{Kervella} et~al.}{2019}]{Kervella2019}
{Kervella} P.,  {Arenou} F.,  {Mignard} F.,   {Th{\'e}venin} F.,  2019, \mn@doi
  [\aap] {10.1051/0004-6361/201834371}, \href
  {https://ui.adsabs.harvard.edu/abs/2019A&A...623A..72K} {623, A72}

\bibitem[\protect\citeauthoryear{{Kim}, {An}, {Stauffer}, {Lee}, {Terndrup}  \&
  {Johnson}}{{Kim} et~al.}{2016}]{Kim2016}
{Kim} B.,  {An} D.,  {Stauffer} J.~R.,  {Lee} Y.~S.,  {Terndrup} D.~M.,
  {Johnson} J.~A.,  2016, \mn@doi [\apjs] {10.3847/0067-0049/222/2/19}, \href
  {https://ui.adsabs.harvard.edu/abs/2016ApJS..222...19K} {222, 19}

\bibitem[\protect\citeauthoryear{{Klose}, {Fuhr}  \& {Wiese}}{{Klose}
  et~al.}{2002}]{Klose2002}
{Klose} J.~Z.,  {Fuhr} J.~R.,   {Wiese} W.~L.,  2002, \mn@doi [Journal of
  Physical and Chemical Reference Data] {10.1063/1.1448482}, \href
  {https://ui.adsabs.harvard.edu/abs/2002JPCRD..31..217K} {31, 217}

\bibitem[\protect\citeauthoryear{{Kovtyukh}, {Soubiran}  \& {Belik}}{{Kovtyukh}
  et~al.}{2004}]{Kovtyukh2004}
{Kovtyukh} V.~V.,  {Soubiran} C.,   {Belik} S.~I.,  2004, \mn@doi [\aap]
  {10.1051/0004-6361:20041449}, \href
  {https://ui.adsabs.harvard.edu/abs/2004A&A...427..933K} {427, 933}

\bibitem[\protect\citeauthoryear{{Kurucz}, {Furenlid}, {Brault}  \&
  {Testerman}}{{Kurucz} et~al.}{1984}]{Kurucz1984}
{Kurucz} R.~L.,  {Furenlid} I.,  {Brault} J.,   {Testerman} L.,  1984, {Solar
  flux atlas from 296 to 1300 nm}

\bibitem[\protect\citeauthoryear{{Lawler}, {Den Hartog}, {Sneden}  \&
  {Cowan}}{{Lawler} et~al.}{2006}]{Lawler2006}
{Lawler} J.~E.,  {Den Hartog} E.~A.,  {Sneden} C.,   {Cowan} J.~J.,  2006,
  \mn@doi [\apjs] {10.1086/498213}, \href
  {https://ui.adsabs.harvard.edu/abs/2006ApJS..162..227L} {162, 227}

\bibitem[\protect\citeauthoryear{{Lawler}, {Sneden}, {Cowan}, {Ivans}  \& {Den
  Hartog}}{{Lawler} et~al.}{2009}]{Lawler2009}
{Lawler} J.~E.,  {Sneden} C.,  {Cowan} J.~J.,  {Ivans} I.~I.,   {Den Hartog}
  E.~A.,  2009, \mn@doi [\apjs] {10.1088/0067-0049/182/1/51}, \href
  {https://ui.adsabs.harvard.edu/abs/2009ApJS..182...51L} {182, 51}

\bibitem[\protect\citeauthoryear{{Lawler}, {Guzman}, {Wood}, {Sneden}  \&
  {Cowan}}{{Lawler} et~al.}{2013}]{Lawler2013}
{Lawler} J.~E.,  {Guzman} A.,  {Wood} M.~P.,  {Sneden} C.,   {Cowan} J.~J.,
  2013, \mn@doi [\apjs] {10.1088/0067-0049/205/2/11}, \href
  {https://ui.adsabs.harvard.edu/abs/2013ApJS..205...11L} {205, 11}

\bibitem[\protect\citeauthoryear{{Lawler}, {Wood}, {Den Hartog}, {Feigenson},
  {Sneden}  \& {Cowan}}{{Lawler} et~al.}{2014}]{Lawler2014}
{Lawler} J.~E.,  {Wood} M.~P.,  {Den Hartog} E.~A.,  {Feigenson} T.,  {Sneden}
  C.,   {Cowan} J.~J.,  2014, \mn@doi [\apjs] {10.1088/0067-0049/215/2/20},
  \href {https://ui.adsabs.harvard.edu/abs/2014ApJS..215...20L} {215, 20}

\bibitem[\protect\citeauthoryear{{Lawler}, {Sneden}  \& {Cowan}}{{Lawler}
  et~al.}{2015}]{Lawler2015}
{Lawler} J.~E.,  {Sneden} C.,   {Cowan} J.~J.,  2015, \mn@doi [\apjs]
  {10.1088/0067-0049/220/1/13}, \href
  {https://ui.adsabs.harvard.edu/abs/2015ApJS..220...13L} {220, 13}

\bibitem[\protect\citeauthoryear{{Lawler}, {Sneden}, {Nave}, {Den Hartog},
  {Emraho{\u{g}}lu}  \& {Cowan}}{{Lawler} et~al.}{2017}]{Lawler2017}
{Lawler} J.~E.,  {Sneden} C.,  {Nave} G.,  {Den Hartog} E.~A.,
  {Emraho{\u{g}}lu} N.,   {Cowan} J.~J.,  2017, \mn@doi [\apjs]
  {10.3847/1538-4365/228/1/10}, \href
  {https://ui.adsabs.harvard.edu/abs/2017ApJS..228...10L} {228, 10}

\bibitem[\protect\citeauthoryear{{Lawler}, {Hala}, {Sneden}, {Nave}, {Wood}  \&
  {Cowan}}{{Lawler} et~al.}{2019}]{Lawler2019}
{Lawler} J.~E.,  {Hala} {Sneden} C.,  {Nave} G.,  {Wood} M.~P.,   {Cowan}
  J.~J.,  2019, \mn@doi [\apjs] {10.3847/1538-4365/ab08ef}, \href
  {https://ui.adsabs.harvard.edu/abs/2019ApJS..241...21L} {241, 21}

\bibitem[\protect\citeauthoryear{{Lee} et~al.,}{{Lee} et~al.}{2011}]{Lee2011}
{Lee} Y.~S.,  et~al., 2011, \mn@doi [\aj] {10.1088/0004-6256/141/3/90}, \href
  {https://ui.adsabs.harvard.edu/abs/2011AJ....141...90L} {141, 90}

\bibitem[\protect\citeauthoryear{{Luck}}{{Luck}}{2017}]{Luck2017}
{Luck} R.~E.,  2017, \mn@doi [\aj] {10.3847/1538-3881/153/1/21}, \href
  {https://ui.adsabs.harvard.edu/abs/2017AJ....153...21L} {153, 21}

\bibitem[\protect\citeauthoryear{{Ludwig}, {Freytag}  \& {Steffen}}{{Ludwig}
  et~al.}{1999}]{Ludwig1999}
{Ludwig} H.-G.,  {Freytag} B.,   {Steffen} M.,  1999, \mn@doi [\aap]
  {10.48550/arXiv.astro-ph/9811179}, \href
  {https://ui.adsabs.harvard.edu/abs/1999A&A...346..111L} {346, 111}

\bibitem[\protect\citeauthoryear{{Magic}, {Weiss}  \& {Asplund}}{{Magic}
  et~al.}{2015}]{Magic2015}
{Magic} Z.,  {Weiss} A.,   {Asplund} M.,  2015, \mn@doi [\aap]
  {10.1051/0004-6361/201423760}, \href
  {https://ui.adsabs.harvard.edu/abs/2015A&A...573A..89M} {573, A89}

\bibitem[\protect\citeauthoryear{{Martell} et~al.,}{{Martell}
  et~al.}{2017}]{martell2017}
{Martell} S.~L.,  et~al., 2017, \mn@doi [\mnras] {10.1093/mnras/stw2835}, \href
  {https://ui.adsabs.harvard.edu/abs/2017MNRAS.465.3203M} {465, 3203}

\bibitem[\protect\citeauthoryear{{Masana}, {Jordi}  \& {Ribas}}{{Masana}
  et~al.}{2006}]{Masana2006}
{Masana} E.,  {Jordi} C.,   {Ribas} I.,  2006, \mn@doi [\aap]
  {10.1051/0004-6361:20054021}, \href
  {https://ui.adsabs.harvard.edu/abs/2006A&A...450..735M} {450, 735}

\bibitem[\protect\citeauthoryear{{Mel{\'e}ndez} \& {Barbuy}}{{Mel{\'e}ndez} \&
  {Barbuy}}{2009}]{Melendez2009}
{Mel{\'e}ndez} J.,  {Barbuy} B.,  2009, \mn@doi [\aap]
  {10.1051/0004-6361/200811508}, \href
  {https://ui.adsabs.harvard.edu/abs/2009A&A...497..611M} {497, 611}

\bibitem[\protect\citeauthoryear{{Mishenina}, {Soubiran}, {Kovtyukh}  \&
  {Korotin}}{{Mishenina} et~al.}{2004}]{Mishenina2004}
{Mishenina} T.~V.,  {Soubiran} C.,  {Kovtyukh} V.~V.,   {Korotin} S.~A.,  2004,
  \mn@doi [\aap] {10.1051/0004-6361:20034454}, \href
  {https://ui.adsabs.harvard.edu/abs/2004A&A...418..551M} {418, 551}

\bibitem[\protect\citeauthoryear{{Mishenina}, {Gorbaneva}, {Basak}, {Soubiran}
  \& {Kovtyukh}}{{Mishenina} et~al.}{2011}]{Mishenina2011}
{Mishenina} T.~V.,  {Gorbaneva} T.~I.,  {Basak} N.~Y.,  {Soubiran} C.,
  {Kovtyukh} V.~V.,  2011, \mn@doi [Astronomy Reports]
  {10.1134/S1063772911080075}, \href
  {https://ui.adsabs.harvard.edu/abs/2011ARep...55..689M} {55, 689}

\bibitem[\protect\citeauthoryear{{Mishenina}, {Pignatari}, {Korotin},
  {Soubiran}, {Charbonnel}, {Thielemann}, {Gorbaneva}  \& {Basak}}{{Mishenina}
  et~al.}{2013}]{Mishenina2013}
{Mishenina} T.~V.,  {Pignatari} M.,  {Korotin} S.~A.,  {Soubiran} C.,
  {Charbonnel} C.,  {Thielemann} F.~K.,  {Gorbaneva} T.~I.,   {Basak} N.~Y.,
  2013, \mn@doi [\aap] {10.1051/0004-6361/201220687}, \href
  {https://ui.adsabs.harvard.edu/abs/2013A&A...552A.128M} {552, A128}

\bibitem[\protect\citeauthoryear{{Mishenina}, {Gorbaneva}, {Pignatari},
  {Thielemann}  \& {Korotin}}{{Mishenina} et~al.}{2015}]{Mishenina2015}
{Mishenina} T.,  {Gorbaneva} T.,  {Pignatari} M.,  {Thielemann} F.~K.,
  {Korotin} S.~A.,  2015, \mn@doi [\mnras] {10.1093/mnras/stv2038}, \href
  {https://ui.adsabs.harvard.edu/abs/2015MNRAS.454.1585M} {454, 1585}

\bibitem[\protect\citeauthoryear{{Molaro} \& {Monai}}{{Molaro} \&
  {Monai}}{2012}]{Molaro2012}
{Molaro} P.,  {Monai} S.,  2012, \mn@doi [\aap] {10.1051/0004-6361/201118675},
  \href {https://ui.adsabs.harvard.edu/abs/2012A&A...544A.125M} {544, A125}

\bibitem[\protect\citeauthoryear{{Moore}, {Minnaert}  \& {Houtgast}}{{Moore}
  et~al.}{1966}]{moore1966}
{Moore} C.~E.,  {Minnaert} M.~G.~J.,   {Houtgast} J.,  1966, {The solar
  spectrum 2935 A to 8770 A}

\bibitem[\protect\citeauthoryear{{Pagel} \& {Patchett}}{{Pagel} \&
  {Patchett}}{1975}]{pagel1975}
{Pagel} B.~E.~J.,  {Patchett} B.~E.,  1975, \mn@doi [\mnras]
  {10.1093/mnras/172.1.13}, \href
  {https://ui.adsabs.harvard.edu/abs/1975MNRAS.172...13P} {172, 13}

\bibitem[\protect\citeauthoryear{{Pehlivan Rhodin}, {Hartman}, {Nilsson}  \&
  {J{\"o}nsson}}{{Pehlivan Rhodin} et~al.}{2017}]{Rhodin2017}
{Pehlivan Rhodin} A.,  {Hartman} H.,  {Nilsson} H.,   {J{\"o}nsson} P.,  2017,
  \mn@doi [\aap] {10.1051/0004-6361/201629849}, \href
  {https://ui.adsabs.harvard.edu/abs/2017A&A...598A.102P} {598, A102}

\bibitem[\protect\citeauthoryear{{Petit}, {Louge}, {Th{\'e}ado}, {Paletou},
  {Manset}, {Morin}, {Marsden}  \& {Jeffers}}{{Petit} et~al.}{2014}]{petit2014}
{Petit} P.,  {Louge} T.,  {Th{\'e}ado} S.,  {Paletou} F.,  {Manset} N.,
  {Morin} J.,  {Marsden} S.~C.,   {Jeffers} S.~V.,  2014, \mn@doi [\pasp]
  {10.1086/676976}, \href
  {https://ui.adsabs.harvard.edu/abs/2014PASP..126..469P} {126, 469}

\bibitem[\protect\citeauthoryear{{Placco} et~al.,}{{Placco}
  et~al.}{2021}]{placco2021}
{Placco} V.~M.,  et~al., 2021, \mn@doi [\apjl] {10.3847/2041-8213/abf93d},
  \href {https://ui.adsabs.harvard.edu/abs/2021ApJ...912L..32P} {912, L32}

\bibitem[\protect\citeauthoryear{{Rebolo}, {Molaro}  \& {Beckman}}{{Rebolo}
  et~al.}{1988}]{Rebolo1988}
{Rebolo} R.,  {Molaro} P.,   {Beckman} J.~E.,  1988, \aap, \href
  {https://ui.adsabs.harvard.edu/abs/1988A&A...192..192R} {192, 192}

\bibitem[\protect\citeauthoryear{{Rice} \& {Brewer}}{{Rice} \&
  {Brewer}}{2020}]{Rice2020}
{Rice} M.,  {Brewer} J.~M.,  2020, \mn@doi [\apj] {10.3847/1538-4357/ab9f96},
  \href {https://ui.adsabs.harvard.edu/abs/2020ApJ...898..119R} {898, 119}

\bibitem[\protect\citeauthoryear{{Shi}, {Gehren}  \& {Zhao}}{{Shi}
  et~al.}{2011}]{shi2011}
{Shi} J.~R.,  {Gehren} T.,   {Zhao} G.,  2011, \mn@doi [\aap]
  {10.1051/0004-6361/201117658}, \href
  {https://ui.adsabs.harvard.edu/abs/2011A&A...534A.103S} {534, A103}

\bibitem[\protect\citeauthoryear{{Sneden}}{{Sneden}}{1973}]{sneden1973}
{Sneden} C.~A.,  1973, PhD thesis, University of Texas, Austin

\bibitem[\protect\citeauthoryear{{Song}, {Alexeeva}, {Sitnova}, {Wang}, {Grupp}
   \& {Zhao}}{{Song} et~al.}{2020}]{Song2020}
{Song} N.,  {Alexeeva} S.,  {Sitnova} T.,  {Wang} L.,  {Grupp} F.,   {Zhao} G.,
   2020, \mn@doi [\aap] {10.1051/0004-6361/201937110}, \href
  {https://ui.adsabs.harvard.edu/abs/2020A&A...635A.176S} {635, A176}

\bibitem[\protect\citeauthoryear{{Soubiran} \& {Girard}}{{Soubiran} \&
  {Girard}}{2005}]{soubiran2005}
{Soubiran} C.,  {Girard} P.,  2005, \mn@doi [\aap]
  {10.1051/0004-6361:20042390}, \href
  {https://ui.adsabs.harvard.edu/abs/2005A&A...438..139S} {438, 139}

\bibitem[\protect\citeauthoryear{{Soubiran}, {Bienaym{\'e}}  \&
  {Siebert}}{{Soubiran} et~al.}{2003}]{soubiran2003}
{Soubiran} C.,  {Bienaym{\'e}} O.,   {Siebert} A.,  2003, \mn@doi [\aap]
  {10.1051/0004-6361:20021615}, \href
  {https://ui.adsabs.harvard.edu/abs/2003A&A...398..141S} {398, 141}

\bibitem[\protect\citeauthoryear{{Sozzetti}, {Torres}, {Latham}, {Stefanik},
  {Korzennik}, {Boss}, {Carney}  \& {Laird}}{{Sozzetti}
  et~al.}{2009}]{Sozzetti2009}
{Sozzetti} A.,  {Torres} G.,  {Latham} D.~W.,  {Stefanik} R.~P.,  {Korzennik}
  S.~G.,  {Boss} A.~P.,  {Carney} B.~W.,   {Laird} J.~B.,  2009, \mn@doi [\apj]
  {10.1088/0004-637X/697/1/544}, \href
  {https://ui.adsabs.harvard.edu/abs/2009ApJ...697..544S} {697, 544}

\bibitem[\protect\citeauthoryear{{Steinmetz} et~al.,}{{Steinmetz}
  et~al.}{2006}]{steinmetz2006}
{Steinmetz} M.,  et~al., 2006, \mn@doi [\aj] {10.1086/506564}, \href
  {https://ui.adsabs.harvard.edu/abs/2006AJ....132.1645S} {132, 1645}

\bibitem[\protect\citeauthoryear{{Takeda}}{{Takeda}}{2023}]{takeda2023}
{Takeda} Y.,  2023, \mn@doi [\actaa] {10.32023/0001-5237/73.1.3}, \href
  {https://ui.adsabs.harvard.edu/abs/2023AcA....73...35T} {73, 35}

\bibitem[\protect\citeauthoryear{{Takeda}, {Zhao}, {Takada-Hidai}, {Chen},
  {Saito}  \& {Zhang}}{{Takeda} et~al.}{2003}]{Takeda2003}
{Takeda} Y.,  {Zhao} G.,  {Takada-Hidai} M.,  {Chen} Y.-Q.,  {Saito} Y.-J.,
  {Zhang} H.-W.,  2003, \mn@doi [\cjaa] {10.1088/1009-9271/3/4/316}, \href
  {https://ui.adsabs.harvard.edu/abs/2003ChJAA...3..316T} {3, 316}

\bibitem[\protect\citeauthoryear{{Takeda}, {Kawanomoto}, {Honda}, {Ando}  \&
  {Sakurai}}{{Takeda} et~al.}{2007}]{Takeda2007}
{Takeda} Y.,  {Kawanomoto} S.,  {Honda} S.,  {Ando} H.,   {Sakurai} T.,  2007,
  \mn@doi [\aap] {10.1051/0004-6361:20077220}, \href
  {https://ui.adsabs.harvard.edu/abs/2007A&A...468..663T} {468, 663}

\bibitem[\protect\citeauthoryear{{Valenti} \& {Fischer}}{{Valenti} \&
  {Fischer}}{2005}]{Valenti2005}
{Valenti} J.~A.,  {Fischer} D.~A.,  2005, \mn@doi [\apjs] {10.1086/430500},
  \href {https://ui.adsabs.harvard.edu/abs/2005ApJS..159..141V} {159, 141}

\bibitem[\protect\citeauthoryear{{Yanny} et~al.,}{{Yanny}
  et~al.}{2009}]{yanny2009}
{Yanny} B.,  et~al., 2009, \mn@doi [\aj] {10.1088/0004-6256/137/5/4377}, \href
  {https://ui.adsabs.harvard.edu/abs/2009AJ....137.4377Y} {137, 4377}

\bibitem[\protect\citeauthoryear{{Zhao}, {Zhao}, {Chu}, {Jing}  \&
  {Deng}}{{Zhao} et~al.}{2012}]{zhao2012}
{Zhao} G.,  {Zhao} Y.-H.,  {Chu} Y.-Q.,  {Jing} Y.-P.,   {Deng} L.-C.,  2012,
  \mn@doi [Research in Astronomy and Astrophysics]
  {10.1088/1674-4527/12/7/002}, \href
  {https://ui.adsabs.harvard.edu/abs/2012RAA....12..723Z} {12, 723}

\bibitem[\protect\citeauthoryear{{{\c{S}}ahin}}{{{\c{S}}ahin}}{2017}]{sahin2017}
{{\c{S}}ahin} T.,  2017, Turkish Journal of Physics, \href
  {https://ui.adsabs.harvard.edu/abs/2017TJPh...41..367S} {41, 367}

\bibitem[\protect\citeauthoryear{{{\c{S}}ahin} \& {Bilir}}{{{\c{S}}ahin} \&
  {Bilir}}{2020}]{sahin2020}
{{\c{S}}ahin} T.,  {Bilir} S.,  2020, \mn@doi [\apj]
  {10.3847/1538-4357/aba2d2}, \href
  {https://ui.adsabs.harvard.edu/abs/2020ApJ...899...41S} {899, 41}

\bibitem[\protect\citeauthoryear{{{\c{S}}ahin} \& {Lambert}}{{{\c{S}}ahin} \&
  {Lambert}}{2009}]{sahin2009}
{{\c{S}}ahin} T.,  {Lambert} D.~L.,  2009, \mn@doi [\mnras]
  {10.1111/j.1365-2966.2009.15251.x}, \href
  {https://ui.adsabs.harvard.edu/abs/2009MNRAS.398.1730S} {398, 1730}

\bibitem[\protect\citeauthoryear{{{\c{S}}ahin}, {Lambert}, {Klochkova}  \&
  {Tavolganskaya}}{{{\c{S}}ahin} et~al.}{2011}]{sahin2011}
{{\c{S}}ahin} T.,  {Lambert} D.~L.,  {Klochkova} V.~G.,   {Tavolganskaya}
  N.~S.,  2011, \mn@doi [\mnras] {10.1111/j.1365-2966.2010.17467.x}, \href
  {https://ui.adsabs.harvard.edu/abs/2011MNRAS.410..612S} {410, 612}

\bibitem[\protect\citeauthoryear{{{\c{S}}ahin}, {Lambert}, {Klochkova}  \&
  {Panchuk}}{{{\c{S}}ahin} et~al.}{2016}]{sahin2016}
{{\c{S}}ahin} T.,  {Lambert} D.~L.,  {Klochkova} V.~G.,   {Panchuk} V.~E.,
  2016, \mn@doi [\mnras] {10.1093/mnras/stw1586}, \href
  {https://ui.adsabs.harvard.edu/abs/2016MNRAS.461.4071S} {461, 4071}

\makeatother
\end{thebibliography}

\bsp	
\label{lastpage}
\end{document}